\documentclass[aps,prb,reprint,amsmath,amssymb,superscriptaddress,showpacs,floatfix,twocolumn,longbibliography]{revtex4-2}
\usepackage{graphicx}
\usepackage{dcolumn}
\usepackage{bm}
\usepackage[colorlinks, allcolors=blue]{hyperref}
\usepackage[usenames, dvipsnames]{color}

\begin{document}

\title{Strong Coupling Theory of Superconductivity and Ferroelectric 
 Quantum Criticality in metallic SrTiO$_3$ }
\author{Sudip Kumar Saha}
\affiliation{Physics Department, Ariel University, Ariel 40700, Israel}
\affiliation{Department of Physics, Bar Ilan University, Ramat Gan 5290002, Israel}
\author{Maria N. Gastiasoro}
\affiliation{Donostia International Physics Center, 20018 Donostia-San Sebastian, Spain}
\author{Jonathan Ruhman}
\affiliation{Department of Physics, Bar Ilan University, Ramat Gan 5290002, Israel}

\author{Avraham Klein}
\affiliation{Physics Department, Ariel University, Ariel 40700, Israel}


\begin{abstract}
Superconductivity in doped SrTiO$_3$ has remained an enduring mystery for over 50 years. The material's status as a ``quantum" ferroelectric metal, characterized by a soft polar mode, suggests that quantum criticality could play a pivotal role in the emergence of its superconducting state.
We show that the system is amenable to a strong coupling (Eliashberg) pairing analysis,
with the dominant coupling to the soft mode being a ``dynamical'' Rashba coupling.  We compute the expected $T_c$ for the entire phase diagram, all the way to the quantum critical point and beyond. We demonstrate that the linear coupling is sufficient to obtain a rough approximation of the experimentally measured phase diagram, but that nonlinear coupling terms are crucial in reproducing the finer features in the ordered phase. The primary role of nonlinear terms at the peak of the superconducting dome is to enhance the effective linear coupling induced by the broken order, shifting the dome's maximum into the ordered phase. Our theory quantitatively reproduces the three-dimensional experimental phase diagram in the space of carrier density, distance from the quantum critical point and temperature, and allows us to estimate microscopic parameters from the experimental data.
\end{abstract}
\maketitle

\section{\label{sec1}Introduction}
Among the family of unconventional superconductors whose behavior has puzzled the scientific community, SrTiO$_3$ (STO) may hold the dubious honor of ``oldest unsolved problem.'' Superconductivity in STO was discovered in 1964 ~\cite{marvin1964}, but despite having quite low $T_c \sim 0.5$ K, it has been poorly understood ever since. The main reasons for this are twofold. First, the typical carrier densities at which superconductivity appear are rather low, with the maximum $T_c$ at approximately $10^{20}$ carriers per cm$^3$ and finite $T_c$ down to $10^{17}/\mbox{cm}^3$ ~\cite{pfeiffer1967,behnia2014,behnia2017, behnia2019}. This makes BCS theory formally invalid as the Debye frequency is higher than $E_F$, implies that $T_c/E_F$ is small but not tiny, and indicates that long-range forces are involved in the pairing ~\cite{jonathan2020}.
The second is that even in its pristine, undoped form, STO is a quantum paraelectric (PE) with a large ($\sim 10^4)$ dielectric constant ~\cite{muller_burkard_1979, sakudo1976, weaver1959} that implies it resides near a quantum critical point (QCP) to a ferroelectric (FE) state~\cite{rowley2014ferroelectric}. 

This makes metallic STO a textbook example of a quantum ferroelectric metal (QFEM), a family of quantum materials whose systems evince soft polar fluctuations alongside gapless electronic excitations. These systems are fascinating in their own right, since the  inversion-breaking fluctuations they host invalidate the usual framework of Fermi liquid theory,  and are always associated with strong spin-orbit coupling~\cite{maria1, maria2} that leads to exotic quantum phenomena. STO can be readily doped or strained across the FE phase transition  ~\cite{tunetaro1976, muller1984, bednorz1994,  tarakanov1996, nakamura_STO_1999, lindner1997, schlom2004, triscone2007, vandermarel2016, behnia_nat_phys2017, rowley2018, tomioka2019, sochnikov2019, harter2019,stemmer2020_2, tomioka2022, tomioka2022}, which persists in the metallic state, making STO an example \emph{par excellence} of a quantum critical superconductor.

A natural pathway to resolve the problems is to use one to fix the other, and there have been a variety of suggestions that pairing in STO is driven by the soft, long-ranged polar fluctuations of the FE mode~\cite{edge2015quantum,wolfle2018superconductivity,maria2}. Crucially, this implies that STO's phenomenology can be at least qualitatively captured by a ``toy'' electron-boson model coupling itinerant electrons to a soft bosonic mode ~\cite{avi_qfem}. Such models have been studied intensively as a description of unconventional superconductivity~\cite{millis1993effect,Son1999,Metzner2003,Chubukov2005a,wang2014charge,oganesyan2001quantum,mross2010controlled,fitzpatrick2013non}. Unfortunately, such attempts run immediately into yet another issue: the polar fluctuations in STO are those of a zone-center \emph{transverse} optical (TO) polar phonon~\cite{kadlec2009,roussev2003theory,yamanaka2000,jonathan2020}. Electrons in polar materials typically couple to the longitudinal mode~\cite{Gurevich1962}, and the gradient coupling to the transverse sector vanishes exactly in the IR limit at the $\Gamma$ point~\cite{ruhman2019comment}. This decoupling is associated with the LO-TO gap, where the zero-$q$ longitudinal mode stiffens due to the Coulomb interaction, which remains present even in the metallic (but still dilute) systems. While some interesting methods to bypass this obstacle have been suggested~\cite{gor2017back,wolfle2018superconductivity,Urazhdin2022}, there are two direct solutions one may consider:  either to conjecture some other, more exotic, form of linear coupling or consider nonlinear processes. 

\begin{figure*}[]
\includegraphics[width=\hsize]{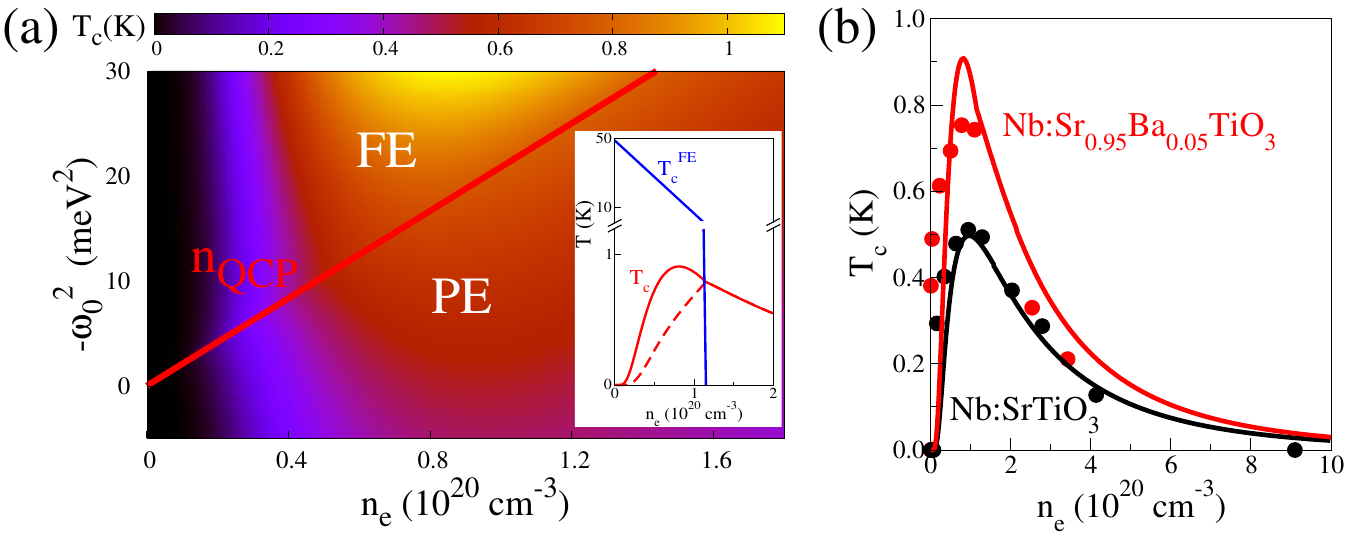}
\caption{\label{dometc_g_lin_quad}
The superconducting phase diagram of SrTiO$_3$. (a) The transition temperature $T_c$ as a function of carrier doping ($n_e$) and the polar phonon frequency squared at zero density, $\omega_0^2$.  
The red line denotes the QCP to the FE state at density $n_e=n_{QCP}$ where the TO phonon becomes massless.
(Inset) Comparing the calculated $T_c$ with (solid) and without (dashed) nonlinear corrections to the pairing vertex. The characteristic cusp, which typically appears in QC fermion-boson models, is softened. The peak shifts into the ordered phase. (b) Comparison of the theoretical and experimental $T_c$ for Ba doped SrTiO$_3$ (from Ref. \onlinecite{tomioka2022}).}
\end{figure*}
One possible mechanism for linear electron-phonon (el-ph) coupling is the ``dynamical'' Rashba spin-orbit coupling, where the polar phonon excitations induce Rashba-like chiral spin currents~\cite{kozii2015odd,ruhman2016superconductivity,kozii2019superconductivity,avi_qfem,maria1,maria2}. Symmetry permits the existence of this coupling, which, in turn, implies the appearance of Rashba spin-orbit coupling in case the phonon displacement gains a finite expectation value; a well known consequence of inversion symmetry breaking.
Furthermore, it couples spin and current, and hence does not generate spontaneous current; time-reversal symmetry is preserved and Bloch's theorem is not violated~\cite{avi_qfem}. The Rashba coupling in STO can be obtained from microscopics through the hopping process between $t_{2g}$ orbitals induced by the polar distortions ~\cite{petersen2000simple,khalsa2013theory,ruhman2016superconductivity,maria1,maria2}. A second, mechanism to achieve an
el-ph interaction
is a quadratic coupling of the electrons to the soft phonons (Ngai's mechanism) ~\cite{ngai1974,vandermarel2019,feigelman_STO_1,volkov2022}. The exchange of two soft phonons leads to a long-range attractive pairing interaction.

%

Both coupling mechanisms have been studied theoretically~\cite{kozii2015odd,wang2016topological,maria1,maria2,vandermarel2019,kozii2019superconductivity,volkov2020multiband,volkov2022,avi_qfem,feigelman_STO_2,feigelman_STO_1,chaudhary2023superconductivity}. In metallic STO without doping towards the FE state (but with carrier doping), the linear coupling yields a phonon-driven attraction that may be analyzed via a BCS-like treatment of a pairing ladder~\cite{maria1,maria2}. The nonlinear interaction gives rise to a non-BCS-like attraction whose strength is given by a direct long-range coupling, that is cut off by the small Fermi surface~\cite{feigelman_STO_1,feigelman_STO_2}. The linear mechanism fails at the lowest carrier densities~\cite{ruhman2016superconductivity}, while the nonlinear mechanism is only valid at those lowest densities~\cite{feigelman_STO_1}.
Both mechanisms (when used beyond their regime of validity) give rise to a superconducting (SC) dome that is roughly consistent with the experiment. However, both treatments fail qualitatively as one approaches the QCP and goes to the strong-coupling regime. Hence, it is not possible to conclusively test theory vs. experiment unless one obtains the entire phase diagram, in \emph{both} carrier density \emph{and} FE order. This requires a self-consistent, strong coupling theory that can fully account for the dynamical fluctuations near the critical point.

This paper demonstrates that soft FE fluctuations drive superconductivity in STO. We show that both linear and nonlinear coupling terms can be captured by a generalized Rashba coupling, yielding a complete phase diagram for Tc as a function of FE doping and carrier density using an Eliashberg framework. Our theory applies across the entire phase diagram when carrier doping isn’t too low, especially at the dome’s peak where the transition shifts from disordered to ordered states.
At higher carrier densities, linear coupling dominates, and nonlinear fluctuations can be ignored. However, linear theory alone misses one key feature: the Tc maximum in STO lies in the ordered phase, shifted from the quantum critical line. This shift is expected and explained by quadratic coupling, which enhances the linear coupling in the ordered state. Our phase diagram aligns quantitatively with experimental data, using reasonable parameter estimates.

We present an overview of our results in Fig.~\ref{dometc_g_lin_quad}. We use $\omega_T$ to denote the transverse optical phonon frequency, which depends on carrier density $n_e$ in the metallic state. The phonon frequency in the insulating state (i.e. $n_e = 0$) is denoted by  $\omega_0^2 \equiv \omega_T^2(n_e = 0)$, such that it is negative in the FE state and positive in the paraelectric (PE) state. Contrary to convention, we use a linear scale for carrier density to highlight the dome’s peak, with logarithmic plots in the Appendix.
In panel (a), the pairing phase diagram is plotted against $n_e$ and $-\omega_0^2$. The red line is the quantum critical (QC) boundary between PE and FE phases (i.e.  $\omega_T(n_e)=0$).  
The inset compares linear and quadratic coupling predictions, illustrating how the nonlinear term softens the dome’s cusp and shifts it into the ordered phase. In panel (b) we compare our theory with experimental data from Ref.~\cite{tomioka2022}.
Our findings show that the Rashba electron-boson theory captures STO’s key features, demonstrating the value of strong-coupling theories for unconventional superconductors.

The paper is structured as follows:
Sec. \ref{sec2} describes our model, and the Eliashberg analysis.
(Sec. \ref{sec2c} explains the extraction of model inputs from experiments.)
Sec. \ref{sec3} presents our numerical solutions, and Sec.\ref{sec4} discusses and compares the effects of the linear and quadratic coupling.
We conclude with a discussion of some broader implications.

\section{Theoretical methods}{\label{sec2}  
\subsection{\label{sec2a}Model for a quantum ferroelectric metal }
The transition to the FE state in STO is a displacive transition associated with a polar phonon mode, which leads to relative displacements of cations (Ti and Sr) and anions (O) inducing dipole moments and finite polarization.
 The long-range Coulomb interaction between dipoles splits these polar modes into the soft TO modes and a longitudinal optical mode at the zone center ($q \rightarrow 0$). At the FE QCP, the longitudinal mode remains gapped while the TO phonon frequency vanishes. Pristine STO is an ``incipient ferroelectric'', i.e. its $\omega_{T}$ softens hugely with decreasing temperature, but eventually levels off and saturates to a small finite value at $T \rightarrow 0$ ~\cite{yamanaka2000}. This is a consequence of strong quantum fluctuations suppressing the FE order.

While the condensation of the TO phonon is one of the major aspects of the displacive ferroelectricity in an insulator, we are interested in the FE state that coexists with metallic states. 
Free charge carriers can be introduced in the pristine or Ba/Ca-doped STO via Nb doping ~\cite{pfeiffer1967,hwang2018, tomioka2022} or oxygen vacancies (SrTiO$_{3-\delta}$)~\cite{behnia2014, behnia_nat_phys2017, vandermarel2016, vandermarel2022}.  There is no macroscopic polarization in the ordered state due to the screening of the electric field by the itinerant electrons. However, there is still a spontaneous breaking of inversion symmetry, which distinguishes the ordered and disordered state. 
The electronic band structure depends greatly on the inversion symmetry, especially in the presence of spin-orbit coupling ~\cite{mattheiss1972,vandermarel2011,karsten2013,jonathan2020,maria1,maria2}. Thus, we anticipate a direct coupling between the ferroelectric order parameter and electrons. A minimal model to describe such systems takes into account itinerant fermions, the TO phonons, and their coupling via an appropriate el-ph interaction. The coupling mechanism also plays a crucial role in mediating the pairing interaction.

In light of the above discussion, we construct the model to describe the QFEM as follows \cite{avi_qfem}. First, we consider a Fermi liquid Lagrangian to describe the itinerant electrons in a metal,   
\begin{equation}
 \begin{aligned}
L_{FL}= \sum_{\bm{k}} \psi_\alpha^\dagger\left(\bm{k}\right)  \left( \partial_\tau +\epsilon\left(\bm{k}\right)  \right) \psi_\alpha \left(\bm{k}\right).
	 \label{eq:FL_lag}
 \end{aligned}
\end{equation}
Here, $\tau$ is imaginary time and $\alpha$ represents the spin indices. For simplicity, we consider a single parabolic band where $\epsilon\left(\bm{k}\right)=\hbar^2\bm{k}^2/{2m}$ describes electron dispersion. It can be linearized near the Fermi level as $\epsilon\left(\bm{k}\right)=v_F(\vert \bm{k} \vert - k_F)$, where $k_F$ and $v_F$ are Fermi wave-vector and Fermi velocity. 
Another important ingredient to construct the QFEM model is the TO phonon which softens to drive the FE transition. This feature prevails in both the metallic and insulating FE materials. The following Lagrangian describes the TO phonon mode.
\begin{equation}
 \begin{aligned}
	L_{\eta}=  D_0^{-1}  \sum_{\bm{q}} \eta_i(\bm{q}) \left[r+\bm{q}^2 a^2 - \left(\frac{a}{c}\right)^2 \partial_\tau^2  \right]  \eta_i(-\bm{q}).
	 \label{eq:phonon_lag}
 \end{aligned}
\end{equation}
Here, $D_0$ carries the unit of the inverse of energy. $\bm{\eta}$ describes the transverse component of a  dimensionless phonon displacement $\bm{u}$ such that  
\begin{equation}
 \begin{aligned}
	\bm{\eta}(\bm{q})=
 \hat{P}(\hat{q})\cdot\bm{u}(\bm{q})\, , \qquad\qquad P_{ij}(\hat{q})=\delta_{ij}-\hat{q}_i \hat{q}_j \, ,
	 \label{eq:projection_define}
 \end{aligned}
\end{equation}
where the operator $P_{ij}$ describes projection on the transverse sector with $\hat{P}$ representing its matrix form.
The lattice constant and the transverse phonon velocity are denoted by $a$ and $c$, respectively. The parameter $r$ represents the distance from the QCP in the undoped sample. In a carrier-doped sample, $r$ is renormalized by the static correction to the phonon mode energy as discussed in the next Section. It depends on the TO phonon frequency $\omega_{T}$ as $r=\left(\omega_{T} a/c\right)^2$. 
As we discuss in Sec. \ref{sec2c}, its dependence on carrier density can be well estimated by a simple linear dependence,
 \begin{equation}
 \begin{aligned}
	\omega_{T}^2  =\omega_0^2  +\gamma_n \, n_e \, ,
	 \label{eq:omega_dependence}
 \end{aligned}
\end{equation}
where $\omega_0$ represents the TO phonon frequency of the insulating sample, which can be modified by doping to the FE order (Ba/Ca doping).

The final crucial ingredient of the model is the el-ph interaction. As we discussed in the Introduction, in our work we assume that $\bm{\eta}$ is coupled to the itinerant electrons via a Rashba coupling, i.e, a vector coupling of spin and charge simultaneously. It is, 
\begin{equation}
 \begin{aligned}
&  \qquad L_{L} =  \\
&\left(\frac{a}{L}\right)^3 g \sum_{\bm{k},\bm{q}} \eta_i(\bm{q}) \psi_\alpha^\dagger\left(\bm{k}+\frac{\bm{q}}{2}\right)  \left( \hat{k} \times \bm{\sigma}_{\alpha \beta} \right)_i \psi_\beta \left(\bm{k}-\frac{\bm{q}}{2}\right).
	 \label{eq:e-ph_coulping_lag}
 \end{aligned}
\end{equation}
Here, $g$ is the el-ph coupling constant with units of energy. $(a/L)^3$ represents the number of unit cells with $L$ denoting the size of the system. $k$ and $q$ are the fermionic and bosonic momenta, respectively. $\bm{\sigma}$ represents the pseudospin in a spin-orbit coupled system.

Before proceeding, we note that STO undergoes an antiferrodistortive transition and has a tetragonal structure at low temperatures. We neglect this for simplicity in the current work, since it does not have a qualitative impact on our results.

\subsection{\label{sec2b}Critical Eliashberg theory}

We now describe the critical Eliashberg theory used to analyze the QFEM model. (The detailed derivations can be found in Ref.~\cite{avi_qfem} and in Appendix~\ref{appA1}.)
We primarily focus on the disordered side of the transition, where there is no ferroelectric order and discuss the generalization to the ordered side when needed.  

We start the analysis by discussing the bosonic and fermionic self-energy corrections. To this end, we denote their four-vector by $q=(q_0,\bm{q})$ and $k=(k_0,\bm{k})$, respectively, where $q_0$ and $k_0$ are Matsubara frequencies and $\bm{q}$ and $\bm{k}$ are momenta. 
The bosonic self-energy after spin summations and projection on the transverse sector is given by
        $ \hat{\Pi}(q) =\Pi(q) \hat{P}(\hat{q})$,
 where $\Pi(q) $ has the following form
 \begin{equation}\label{eq:bubble_3D_full_mn} 
    \begin{aligned}
    \Pi(q)  &= \delta r- \delta\Pi(q)\,.
    \end{aligned}
 \end{equation}  
Here,  
 \begin{equation}\label{eq:bubble_3D_full_mn1}     
     \begin{aligned}
     \delta r = \bar{g}\nu_F\,,
     \end{aligned}
 \end{equation}  
is the induced shift of the critical point, $r\rightarrow r-\delta r$, and
\begin{equation} 
\bar{g}=g^2 D_0\label{eq:gbar_def}
\end{equation} 
is the effective fermion-boson coupling, where the bare coupling $g$ was defined above in Eq.~\eqref{eq:e-ph_coulping_lag}, and   $\nu_F=k_F^2a^3/2\pi^2 v_F$ is the 3D density of states per spin at the Fermi level. $v_F$ is the Fermi velocity. 

The second term in Eq.~\eqref{eq:bubble_3D_full_mn} represents the dynamical contribution
\begin{equation}
     \begin{aligned}
    \delta\Pi(q) &=\bar{g}\nu_F  \left(\frac{\vert q_0 \vert}{v_F \vert \bm{q} \vert} \right) \arctan{\left(\frac{v_F \vert \bm{q} \vert }{\vert q_0 \vert }\right)}\,,  \\
    \label{eq:bubble_3D_full_mn2}        
    \end{aligned}
\end{equation} 
which causes the bosonic mode to become overdamped at the QCP.
~\cite{chubukov2005,avi_qfem}

Similarly, we can obtain the fermionic self-energy $\Sigma(k_0)$, which involves a logarithmic divergence giving rise to marginal Fermi liquid behavior~\cite{chubukov2005,avi_qfem},
\begin{equation}
 \begin{aligned}
         \Sigma(k_0) &=-i\frac{\bar{g} }{v_F k_a} \frac{1}{4\pi^2} k_0 \ln \left(\frac{\omega_\Lambda}{ \vert k_0 \vert }\right)^{1/3}\,.
          \label{eq:fermion_self_energy_QCP_mn}
 \end{aligned}
\end{equation}
 Here, $\omega_\Lambda$ is an UV cutoff (see Appendix for an exact expression), and $k_a=1/a$.

 Next, we focus on pairing instabilities via a linearized gap equation, which takes the form 
\begin{equation}
 \begin{aligned}
	 \Phi_{\alpha \beta}(k) =\frac{\bar{g} T}{ k_a^3}    \sum_{p_0}\int & \frac{d^3p}{(2\pi)^3}  G(p)G(-p)D(p-k)  \\
	 & \times    \bm{\gamma}^{ \nu \alpha}(-\hat{k})\Phi_{\nu \mu}(k)  \cdot \hat{P}(\hat{q})\cdot \bm{\gamma}^{ \mu \beta}(\hat{k}).
 \label{eq:gapeq_vector1_mn}
 \end{aligned}
\end{equation}
Here, $G$ and $D$ are fermionic and bosonic propagators, respectively.  $\bm{\gamma}(\hat{k})=\hat{k} \times \bm{\sigma}$ is the interaction form factor that arises due to the vector coupling in Eq.~\eqref{eq:e-ph_coulping_lag}.

Since boson modes are slow compared to fermions, the maximum scattering is favored when the momentum transfer $q$ is parallel to the Fermi surface ~\cite{chubukov2005,avi_qfem}. 
This assumption allows us to factorize the momentum integration in both the expression of $\Sigma$ and the gap equation and yields an effective frequency dependent gap equation. The frequency dependent pairing attraction $d(q_0)$  is,
  \begin{equation}
 \begin{aligned}
      d(q_0) =\int_0^{\Lambda} q_\parallel dq_\parallel D_{q_\parallel}  
     =\ln\left[ 1+ \frac{\omega_\Lambda}{\vert q_0 \vert} \right]^{1/3},
          \label{eq:boson_function_QCP_mn}
 \end{aligned}
\end{equation}
which is obtained by integrating the bosonic propagator over the momentum component $q_\parallel$ parallel to the Fermi surface and taken in the limit $q_0 \ll  v_F \vert \bm{q} \vert$.

Considering only the singlet pairing, which is the leading instability~\cite{avi_qfem, maria1}, summing over spin  and projecting on the transverse sector,  Eq.~\eqref{eq:gapeq_vector1_mn} can be further simplified as follows
 \begin{equation}
 \begin{aligned}
	 \phi(k_0)  &=  \lambda T \sum_{p_0} \frac{ d(p_0-k_0)\phi(p_0)}{\vert \tilde\Sigma(p_0)\vert}\,.
 \label{eq:gapeq_vector2_mn}
 \end{aligned}
\end{equation}
Here, we have defined $\tilde{\Sigma}(p_0)=p_0+\Sigma(p_0)$. The dimensionless coupling constant $\lambda$ will be discussed in detail in the next section. The above equation leads to the expression for $T_c$ at the QCP ~\cite{chubukov2005,avi_qfem} 
\begin{equation}
 \begin{aligned}
	 T_c \sim \omega_\Lambda \exp\left( -\pi^2 \sqrt{\frac{3v_Fk_a}{\bar{g}} }\right),
 \end{aligned}
\end{equation}
where the dependence on $\sqrt{ \bar{ g}}$ is a direct consequence of the logarithmic divergence and, thus, $T_c$ is enhanced compared to BCS theory ($T_c^{\text{BCS}} \propto \exp(-1/\bar{ g})$).

\begin{figure*}[t]
\includegraphics[width=\hsize]{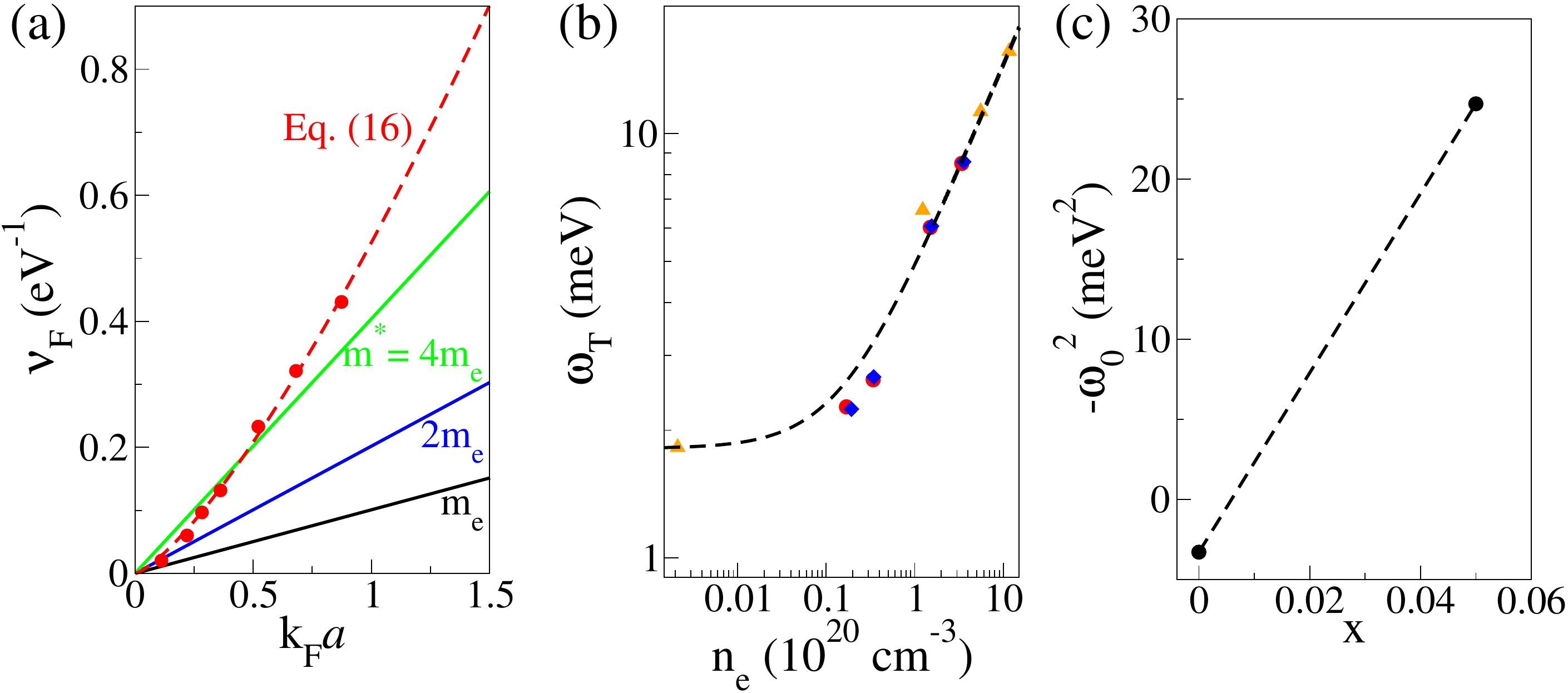}
\caption{\label{fig_dos_omega} Material parameters of STO. (a) The density of states per spin as a function of density $k_Fa$. 
The solid lines denote $\nu_F=m^*k_Fa^3/2\pi^2 \hbar^2$, when $m^*$ is considered to be equal to once, twice and four times the free electron mass $m_e$. The red dashed line demonstrates Eq.~\eqref{eq:ndos_fit} which fits the data from specific heat measurements (red circles). It indicates $m^* > 4 m_e$ for carrier density $n_e > 10^{20}$ cm$^{-3}$
(b)The TO phonon frequency for STO without FE doping is shown as a function of carrier density. The red circles, blue diamonds, and orange triangles represent the measured data from Ref.~\onlinecite{mazin2008},  Ref.~\onlinecite{devreese2010}, and Ref.~\onlinecite{yoon2021}, respectively. The black dashed line (Eq.~\eqref{eq:omega_dependence} with $\omega_0^2=3.3$ meV$^2$ and $\gamma_n \approx 2.1 \times 10^{-19}$ meV$^2$ cm$^{3}$) fits the experimental data sets. 
(c)   $\omega_0^2$ is plotted as a function of Ba content $x$ in Sr$_{1-x}$Ba$_x$TiO$_3$. The black dashed line corresponds to Eq.~\eqref{eq:omega_param}. }
\end{figure*}

Eq. \eqref{eq:gapeq_vector2_mn}, when treated as an eigenvalue problem, can be solved numerically exactly for $T_c$ in the entire phase diagram of $n_e, \omega_0^2$, provided two technical issues are dealt with. First, the $q_0=0$ divergence of $d_0$ in Eq. \eqref{eq:boson_function_QCP_mn} must be dealt with. This can be done ~\cite{sachdev1988,chubukov2008,moon2010, chubukov2020} by converting from the pairing function to the gap, 
$\Delta(k_0)=k_0 \, \phi(k_0)/\tilde{\Sigma}(k_0)$ where $\tilde{\Sigma}(k_0)=k_0+\Sigma(k_0)$ (see Appendix~\ref{appA3} for details and the resulting equation). Second, to account for the ordered phase one should modify the bosonic and fermionic bare actions. In this work, for simplicity, we assume that we are close enough to the QCP that we can (a) neglect the feedback of the order on the fermionic propagators and (b) describe the ordered bosonic state using a simple Ginzburg-Landau quadratic action. These assumptions are enough to generate all the results presented in our work.

\subsection{\label{sec2c} Extraction of parameters from experimental data }
In the previous section, we presented the Eliashberg formalism leading to the gap equation which we use to compute $T_c$, Eq. \eqref{eq:gapeq_vector2_mn}.  The two key parameters in this equation are the effective coupling constant, $\lambda$, and the transverse optical (TO) phonon frequency, $\omega_T$ in Eq.~\eqref{eq:omega_dependence}, which controlls the distance from the QCP. In this section, we discuss their values in light of experiments and \emph{ab initio} calculations.

We start with the coupling $\lambda$. It depends on the Rashba coupling and density of states as follows,
\begin{equation}
 \begin{aligned}
     \lambda= g_{TO}^2 D_0 \nu_F\, ,
	 \label{eq:param_lambda}
 \end{aligned}
\end{equation}
where 
\begin{equation}
    g_{TO} = \frac{g}{2 k_F a}
\end{equation}
and $D_0$ was defined above, see Eq. \eqref{eq:phonon_lag} (in our definition  $D_0^{-1} = 1$ meV ~\cite{maria2}).  
The fermionic DOS, $\nu_F$, is extracted from specific heat measurements~\cite{mccalla2019}. Empirically, it exhibits a power law dependence on the Fermi momentum 
\begin{equation}
 \begin{aligned}
	\nu_F \simeq A \left( k_F a \right)^{4/3}\,,
	 \label{eq:ndos_fit}
 \end{aligned}
\end{equation}
where $A = 0.52$ eV$^{-1}$ and the lattice constant for STO is $a=3.9$ \AA.
Panel (a) of Fig.~\ref{fig_dos_omega} shows a comparison between the above expression and the measured $\nu_F$  as a function of carrier density (the red-filled circles). As can be seen, 
$\nu_F$ reveals the rise of the effective carrier mass $m^*$ with growing carrier concentration. 

\begin{figure}[b]
\includegraphics[width=\hsize]{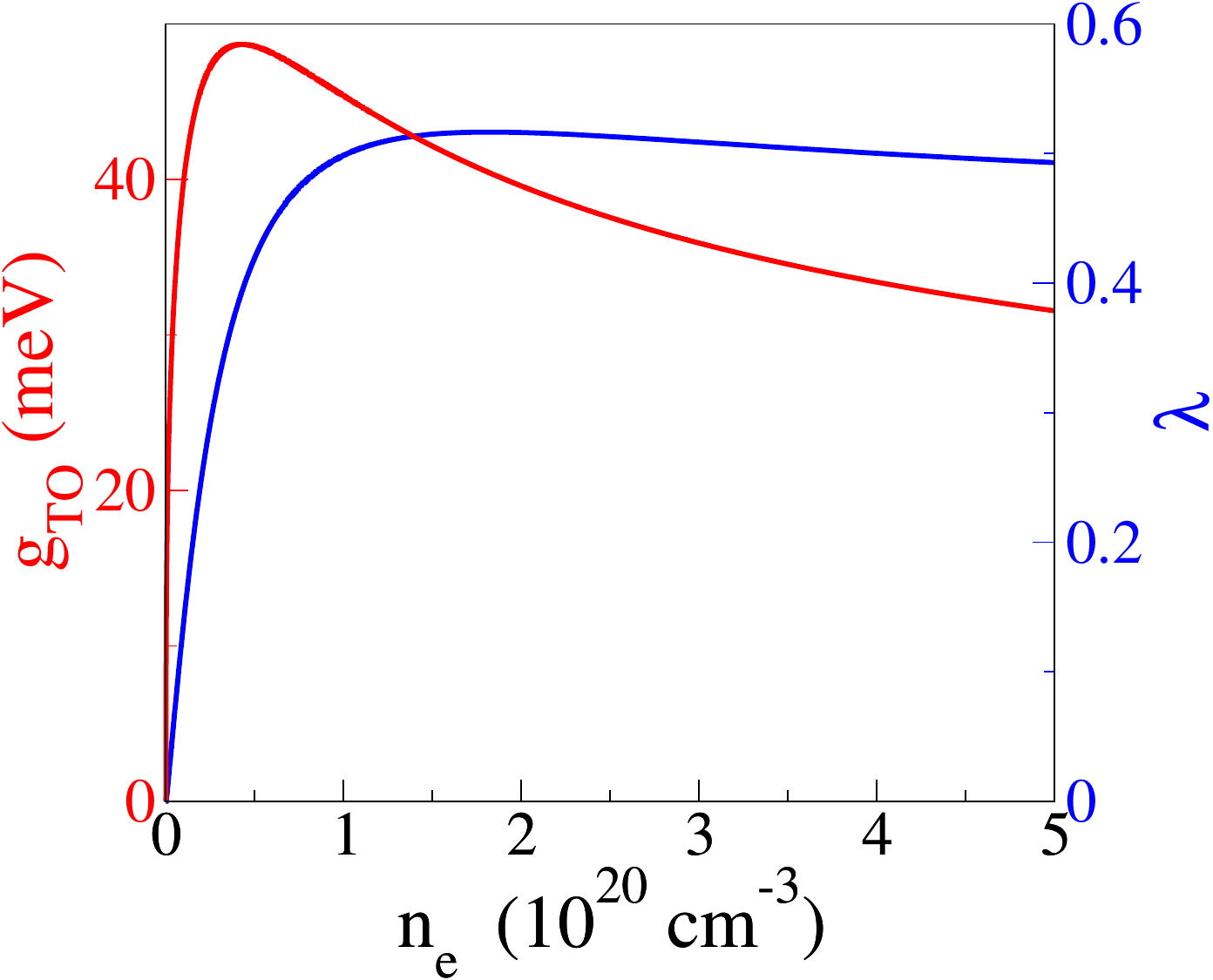}
\caption{\label{fig_gto_DFT} The red solid line represents the bare electron-phonon coupling $g_{TO}$ (left y-axis) in the unit of meV ~\cite{maria2} and the blue solid line denotes the dimensionless effective coupling constant $\lambda$  (right y-axis) as a function of carrier density in pristine STO.}
\end{figure}

Next, we turn to the coupling $g_{TO}$, which was estimated in Ref.~\cite{maria2} using \emph{ab initio} calculations of the band-structure in the presence of an optical distortion. It is extracted from the splitting that rises linearly with momentum $k$ near the $\Gamma$-point. However, this splitting eventually deviates from the linear behavior, reaching a maximum somewhere in the Brillouin zone when $k$ exceeds a small fraction of $1/a$, 
afterwards it decays as $\sim 1/k$. Consequently, the coupling of electrons on the Fermi surface, depends strongly on density~\footnote{According to Ref.~\onlinecite{maria2}, the origin of this 
momentum dependence is the quenching of orbital angular momentum due to a competition between SOC and hopping energies.}. 
Overall, $g_{TO}$ exhibits a dome-like dependence on density, and at its maximum (around at density $n_e=0.43\times 10^{20}$ cm$^{-3}$) this coupling reaches $48.68$ meV. The functional form of this asymptotic behavior is shown in Appendix~\ref{app2}.

In Fig.~\ref{fig_gto_DFT} we plot the coupling, $g_{TO}$, (red solid line; left $y$-axis) and the resulting coupling constant, $\lambda$ in Eq.~\eqref{eq:param_lambda}, (blue solid line; right $y$-axis) as a function of electron density, $n_e$. 
At low-density both couplings increase with $n_e$, while in the high-density regime, the increase in DOS compensates the decrease of $g_{TO}$ making $\lambda$ only weakly density dependent. Interestingly, we found that the \emph{ab initio} value \emph{over}estimates $T_c$ even for STO with no doping towards the FE QCP. Accordingly, $g_{TO}$ in the numerics is taken from Ref.~\onlinecite{maria2} but is scaled down by a factor of order 1 ($g_{TO}\rightarrow g_{TO}/1.1$) to match the maximum $T_c$ of STO without FE doping. This aesthetic change does not of course impact our qualitative results. It is also worth commenting that in the tetragonal phase of STO, the coupling in Eq. \eqref{eq:e-ph_coulping_lag} splits into 3 channels, only one of which contributes to the pairing. We took the largest in our estimates.

\begin{figure*}[t]
\includegraphics[width=\hsize]{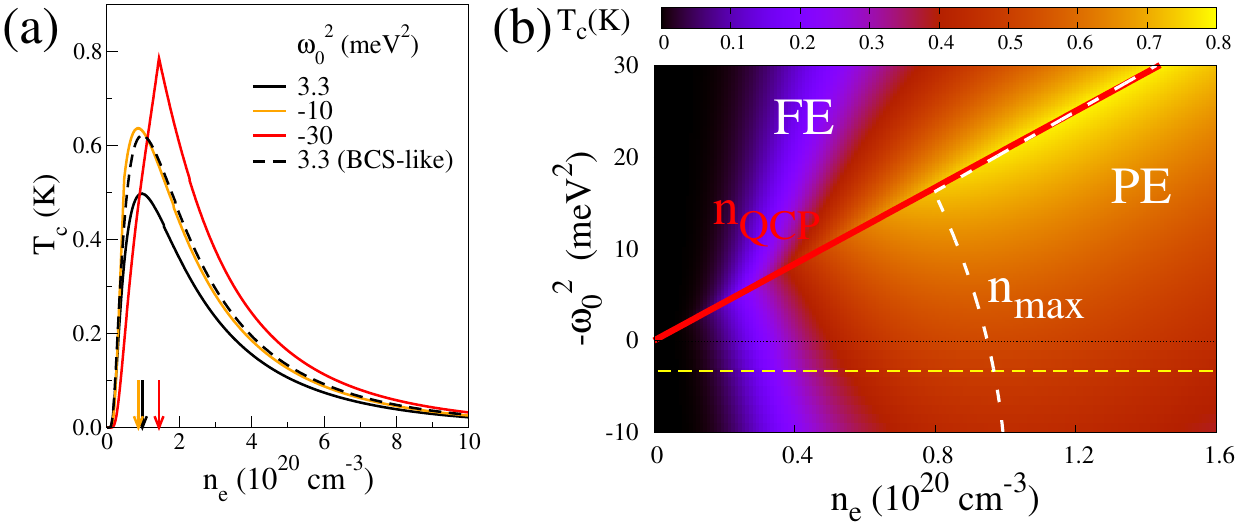}
\caption{\label{dometc_g_DFT} The Eliashberg phase diagram of STO for the linear coupling. (a) $T_c$ as a function of carrier density for various values of  $\omega_0^2$ in unit of meV$^2$. See text for a detailed discussion of coupling constants, etc. 
The arrows mark the peak position of the SC dome, $n_{max}$. 
The black dashed line represents BCS-like $T_c$, calculated without accounting for self-energy corrections.
 (b)  $T_c$ as a function of carrier density along $x$-axis and $-\omega_0^2$ along $y$-axis. The coupling constant is the same as in the left panel. The yellow dashed line marks STO without doping to the FE state ($\omega_{0}^2=3.3$ meV$^2$). The red solid line and the white dashed line denote $n_{QCP}$ and $n_{max}$, respectively.  
 }
\end{figure*}

Next, we focus on another key parameter; the TO phonon frequency, $\omega_{T}$, which is related to the distance, $r$, from the QCP.
Measurements of the zero-doping value, $\omega_{0}$, show an incomplete softening at low temperatures, signaling a crossover from classical  to quantum paraelectricity ~\cite{muller_burkard_1979,yamanaka2000}. Panel (b) of Fig.~\ref{fig_dos_omega} exhibits the density dependence of $\omega_{T}$, which is captured by the parameter $\gamma_n$ in Eq.~\eqref{eq:omega_dependence}. The filled symbols represent the experimental measurements ~\cite{mazin2008, devreese2010, yoon2021} showing that $\omega_{T}^2$ grows linearly with increasing carrier density $n_e$~\footnote{There exists a recent measurement~\cite{yoon2021} of $\omega_{T}$  at extremely low density ($n_e \sim 2 \times 10^{17}$ cm$^{-3}$), which leads to an accurate extrapolation to $n_e=0$ (where $\omega_{0}^2=3.28$ meV$^2$).}.
By fitting to the data points in panel (b) we obtain $\gamma_n \approx 2.1 \times 10^{-19}$ meV$^2$ cm$^{3}$ ~\cite{kraxenberger1980,jonathan2020,volkov2022, chung2022}.

Ba/Ca doping or strain pushes STO through the QCP which is reflected in the phonon frequency $\omega_0$ in Eq.~\eqref{eq:omega_dependence}. In this work, we compare to data on Ba doping, and we assume a simple relation,
 \begin{equation}
 \begin{aligned}
\omega_0^2 (\text{meV}^2)=\alpha(x_c-x) \, .
 \label{eq:omega_param}
  \end{aligned}
 \end{equation}
 where $x$ denotes the doping level as in the formula
 Sr$_{1-x}$Ba$_x$TiO$_3$.
 Here, $\alpha=560$ meV$^2$ and $\omega_0$ vanishes at the critical doping content $x_c=0.0059$. The function in Eq.~\eqref{eq:omega_param} is also shown by the black dashed line in panel (c) of Fig.~\ref{fig_dos_omega}, where the two data points correspond to insulating pristine STO ($\omega_0^2=3.3$ meV$^2$) and insulating Sr$_{0.95}$Ba$_{0.05}$TiO$_3$ ~\cite{tomioka2022}($\omega_0^2=-24.69$ meV$^2$).

\section{\label{sec3} The theoretical superconducting phase diagram}
In this section, we compute the transition temperature in the space of electronic density and distance from the critical point by solving the gap equation (see Appendix~\ref{appA3}), using the parameters outlined in the previous section. 
As described in Sec.~\ref{sec2b}, our strong coupling theory allows us to calculate $T_c$ all the way to the QCP. In what follows, we discuss the results of this analysis.

The calculated $T_c$ is shown in  Fig.~\ref{dometc_g_DFT}. Panel (a) exhibits the $T_c$ dome as a function of carrier density both for $\omega_0^2 > 0$ (black line) and  $\omega_0^2<0$ (blue and red lines). 
When $\omega_0^2>0$ the system remains in the disordered phase for all values of density. 
In this regime, we also obtain the BCS-like result, depicted as the black dashed line, calculated from the gap equation without incorporating bosonic and fermionic self-energy corrections.
For $\omega_0^2 <0$ the insulating system is in the FE phase and therefore the critical point is approached at some critical density, $n_{QCP}$. As can be seen the existence of such a critical point enhances the maximum $T_c$, causes the dome to develop a cusp and shifts the location of the maximum $T_c$, $n_{max}$. This shift depends on the value of $\omega_0^2$. When it is positive, or negative but close to zero,  $n_{max}$ diminishes with $\omega_0^2$. However, at a certain value of $\omega_0^2$, $n_{max} = n_{QCP}$ and  the location of the maximum becomes slaved to the location of the quantum critical point upon reducing $\omega_0^2$ further.

This is better observed in panel (b) of  Fig.~\ref{dometc_g_DFT}, where we plot a colormap of $T_c$ in the space of the distance from the critical point, $\omega_0^2$, and the electron density $n_e$. We indicate the density dependent critical point, $n_{QCP}$, by the red solid line and the carrier density where  $T_c$ is maximal, $n_{max}$, using the white dashed line. The merging of the two densities is observed at some negative value of $\omega_0^2$.

The slaving of $n_{max}$ to $n_{QCP}$ is directly linked to the behavior of the effective coupling $\lambda$.  Above a density $n_e \sim 10^{20}$ cm$^{-3}$ the coupling becomes density independent as indicated in  Fig.~\ref{fig_gto_DFT}.  Thus, when the critical  density $n_{QCP}$ is of the order of, or larger than, this value  the peak will be slaved to $n_{QCP}$. This is because the critical theory favors the maximum $T_c$ when the critical FE fluctuation is the strongest, i.e., at $r=\omega^2_{TO} a^2/c^2=0$. On the other hand, when $n_{QCP}$ is smaller than this scale, $\lambda$ is strongly density dependent. This causes a competition between the critical enhancement at low density and the growth of the coupling with increasing density, which results in the peak appearing at carrier densities higher than the critical density, $n_{max}>n_{QCP}$.

The above argument can be verified by introducing a density cutoff to the Rashba coupling $g_{TO}$ in the high-density regime. The reason behind this approach is rather intuitive. A sharper fall of coupling, $g_{TO}$, also means quicker decay of the effective coupling, $\lambda$, in the high-density regime. This results in a shift of the maximum $T_c$ into the ordered phase away from the QCP (i.e. $n_{max}< n_{QCP}$ for all $\omega_0^2$ except for a point). In fact, the existence of such a cutoff, compared to the DFT values used in our work, is to be expected, since the DFT calculations neglect screening effects that are expected to reduce SOC at higher densities~\cite{chung2022}. The technical details of this approach and the corresponding phase diagram of $T_c$ are discussed in Appendix~\ref{app4b}.

Unfortunately, experiments in FE Ba/Ca-doped SrTiO$_3$ indeed find that $n_{max}$ is generally smaller than $n_{QCP}$~\cite{tomioka2022}. In addition, the theoretical phase diagram is characterized by a line of cusps in $T_c$. These cusps are a universal feature of QC Eliashberg theory, which are typically ignored when studying these theories, and are never really seen in experiments ~\cite{chubukov2005, avi_qfem}.  Thus, the critical Eliashberg theory with the coupling $\lambda$ in Fig.~\ref{fig_gto_DFT}, while reproducing the general energy scales well, disagrees with experiments at the qualitative level. Even introducing a cutoff into the coupling as discussed above does not reconstruct a reasonable phase diagram, as can be seen in Appendix~\ref{app4b}.
Thus, the key remaining question is whether a feasible physical mechanism exists that can reproduce the finer features of the phase diagram.

\section{\label{sec4} Corrections to the phase diagram from non-linear terms}
In the previous section, we concluded that the QC Eliashberg theory with linear coupling fails to account for some of the qualitative features seen in the experimental SC phase diagram for Ba/Ca doped STO.
A possible origin for the disagreement between our theory and experiment is the omission of non-linear corrections to the electron-phonon coupling. Such terms correspond to the exchange of two TO phonons and, in the Rashba channel, can be described by a correction to the Lagrangian,
\begin{equation}
 \begin{aligned}
L_{NL} &=g_{NL}\left(\frac{a}{L}\right)^3  \sum_{\bm{k},\bm{q}}  \psi_\alpha^\dagger\left(\hat{k}+\frac{\bm{q}}{2}\right) \left[ \bm{\eta}(\bm{q})  \cdot    \left(    \hat{k} \times \bm{\sigma}_{\alpha \gamma}    \right) \right] \\
&  \quad \left[  \bm{\eta}(\bm{q})  \cdot  \left(    \hat{k} \times \bm{\sigma}_{ \gamma \beta}    \right) \right]          \psi_\beta \left(\bm{k}-\frac{\bm{q}}{2}\right) \, .
	 \label{eq:e-ph_vec_quad_coulping_lag}
 \end{aligned}
\end{equation}
Here, $g_{NL}$ represents the bare coupling constant and the transverse lattice displacement vector is denoted by $\bm{\eta}$. We note in passing that unlike the linear coupling to the TO mode, there are no symmetry constraints on the existence of 
a quadratic coupling, as pointed out long ago by Ngai ~\cite{ngai1974}, in a slightly different context. Later this idea was extended to the TO mode, and is currently considered as a possible origin for the low-density superconductivity in STO, even without the linear coupling~\cite{vandermarel2019, feigelman_STO_1, volkov2022}. Accordingly, Eq. \eqref{eq:e-ph_vec_quad_coulping_lag} should be understood as a projection onto the transverse channel of a more general coupling (see Appendix~\ref{app3c}). This should not qualitatively change its impact.

The nonlinear coupling has two important impacts. First, it can mediate pairing via a nonlinear fluctuation contribution, even in the absence of linear pairing. This contribution may be the origin for the low-density superconductivity in STO ~\cite{vandermarel2019, feigelman_STO_1, volkov2022}. It arises from the logarithmic divergence of the soft phonon collective modes, see Eq. \eqref{eq:boson_function_QCP_mn}, which at low densities is cut off only by the small Fermi surface. However, at the higher densities, when the Fermi surface is large, this contribution is washed out and is parametrically small compared to the linear coupling (see Appendix~\ref{app3a}), such that
\begin{equation}
    \frac{d_{NL}}{d} \sim  \left(\ln\frac{v_F k_a}{2\pi T_c}\right)^{-1}
\end{equation}
where $d_{NL}$ is the nonlinear contribution to the pairing kernel, see Eq. \eqref{eq:boson_function_QCP_mn}. Numerically we find the ratio to be on the order of $10^{-3}$ near the top of the dome.

\begin{figure*}[t]
\includegraphics[width=\hsize]{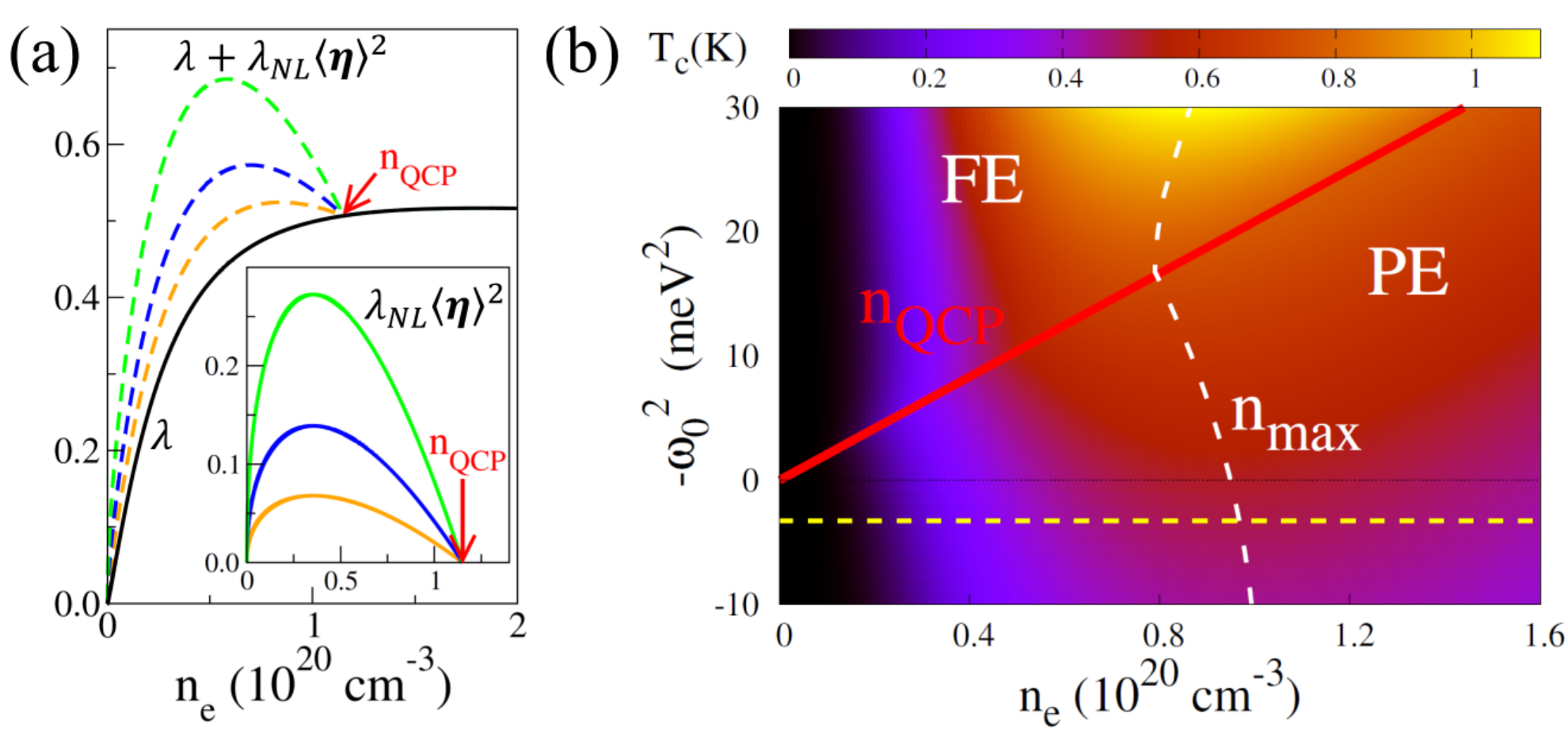}
\caption{\label{quad_lambda_Tc}   The SC phase diagram with nonlinear corrections included. 
(a) The effective coupling constants as a function of the carrier density $n_e$. The black solid line represents the linear coupling, $\lambda$. The inset shows the quadratic coupling induced by the FE order in our model, with
$\omega_0^2=-24$ meV$^2$. The 
the green, blue, and red solid lines correspond to a quartic coefficient $b=25$, $49$, and $100$ meV$^2$, respectively. The corresponding net coupling, $\lambda+\lambda_{NL} \langle \bm{\eta} \rangle^2$, is shown in the main figure by dashed lines with the same color code. The red arrow in both the main figure and inset marks $n_e=n_{QCP}$. 
(b) The color gradient represents $T_c$, calculated at $b=49$ meV$^2$, as a function of $n_e$ and $-\omega_0^2$ along $x$-axis and $y$-axis, respectively. The yellow dashed horizontal line marks STO without FE doping ($\omega_{0}^2=3.3$ meV$^2$). The red solid line denotes $n_{QCP}$ and the white dashed line shows the peak position of the SC dome, $n_e=n_{max}$.}
\end{figure*}

On the other hand, the second contribution can be very relevant even near the top of the dome. Namely, once entering the FE ordered phase the phonon displacement develops a finite expectation value 
$\bm{\eta} = \langle\bm{\eta}\rangle + \delta\bm{\eta}$ with $ \langle\bm{\eta}\rangle \ne 0 $. Plugging this in Eq.~\eqref{eq:e-ph_vec_quad_coulping_lag} one obtains a correction to the \emph{linear} coupling which adds to the existing one. It leads to a gap equation that adopts a form analogous to Eq.~\eqref{eq:gapeq_vector2_mn}, with the effective coupling modified to
\begin{equation}
    \lambda \rightarrow \lambda+\lambda_{NL} \langle \bm{\eta} \rangle^2
\end{equation} 
(see Appendix~\ref{app3c} for details). Here,  
$\lambda_{NL}  =2\bar{g}_{NL} \nu_F /(k_F a)^2$ and $\lambda$ was defined above, see Eq. \eqref{eq:param_lambda}. A comparison of the values of $g_{NL}$ extracted from experiments ~\cite{volkov2022,pfeiffer1967,behnia2017} to the linear coupling yields an estimate $g_{NL} \approx 0.13 g$, and a pairing coupling constant
\begin{equation}
 \begin{aligned}
    \lambda_{NL} = 0.14 (g_{TO}^{m})^2 D_0 \nu_F  \, ,
 \label{eq:param_lambda_quad}
 \end{aligned}
\end{equation}
where the maximum value of the linear coupling, $g_{TO}$, is $g_{TO}^m=48.68$ meV (at $n_e=0.43\times 10^{20}$ cm$^{-3}$).

The contribution of the non-linear coupling is also affected by the mean dimensionless displacement $\langle \bm{\eta}\rangle$. To estimate its value we employ a  free energy analysis $F=-(r/2) |\bm{\eta}|^2+(b/4) |\bm{\eta}|^4$, which is  minimized at $\langle \bm{\eta} \rangle^2=(r/b)$. The value of $b$ can be estimated using experimental data~\cite{kadlec2009}, although in practice we did the calculation for several values of $b$.   
For example in the case of $n_e = 1 \times 10^{20}$ cm$^{-3}$, $\omega_0^2 = -24$ meV$^2$ and $b = 49$ meV$^2$, the maximal displacement can be estimated at $ \bm{\eta}_{max}^2 \sim 0.06$ . 

In panel (a) of Fig.~\ref{quad_lambda_Tc}, we present the nonlinear coupling, $\lambda_{NL} \langle \bm{\eta} \rangle^2$, in the inset, while the linear coupling, $\lambda$, and the net coupling, $\lambda + \lambda_{NL} \langle \bm{\eta} \rangle^2$, are shown in the main figure as functions of density, $n_e$, at $\omega_0^2 = -24$ meV$^2$ for three different values of $b$ ($=25$, 49, and 100 meV$^2$). The nonlinear coupling increases with $n_e$ at low $n_e$, reaching a maximum. This behavior is driven by the increasing density of states. 
At higher $n_e$, the decreasing amplitude of $r = \omega_{T}^2 a^2 / c^2$ outweighs the increase in $\nu_F$, causing the nonlinear coupling to decay with $n_e$ and vanish at $n_{QCP}$, as indicated by the red arrow.
The net coupling rises with $n_e$ at low $n_e$. However, in the regime near QCP within the FE order, the density independent nature of the linear coupling breaks down due to the enhancement from the nonlinear coupling. As a result, the net coupling decays rapidly with $n_e$ in this regime.

We solved the gap equation associated with the generalized Rashba coupling including nonlinear terms to calculate $T_c$ (see Appendix~\ref{app3c} for  details). The resulting phase diagram in the space of $\omega_0^2$ and $n_e$ is shown in panel (b) of Fig.~\ref{quad_lambda_Tc}, with the color gradient representing the value of $T_c$. Notably, the density at which $T_c$ reaches its maximum, $n_e=n_{max}$ (indicated by the white dashed line), decreases as $\omega_0^2$ diminishes when $\omega_0^2$ is small (whether positive or negative). At a certain value of $\omega_0^2$ ($<0$), $n_{max}$ crosses the quantum critical line, $n_{QCP}=n_{max}$ (represented by the red solid line). As $\omega_0^2$ decreases further, $n_{max}$ increases but remains within the ordered phase, satisfying $n_{max}<n_{QCP}$, which is consistent with experimental results.

\section{Discussion}
In this work, we successfully reconstructed the superconducting phase diagram of STO by solving the Eliashberg equations for a QFEM in the strong coupling regime. While the idea that superconductivity in STO is due to FE fluctuations has been around for a while~\cite{edge2015quantum}, we provide a rigorous test of a comprehensive theory, since we provide a 3D phase diagram in temperature, carrier density, and proximity to the FE phase transition in both the ordered and disordered state. Our analysis reproduced correctly both the energy scales and many of the finer details of the STO phase diagram, using parameters taken
only from experimental estimates and \emph{ab initio} calculations (albeit with some uncertainty \footnote{While we did fine-tune some of the values that are less constrained experimentally for aesthetic considerations (e.g. the coupling constants, see Secs. \ref{sec2} and \ref{sec3}), the overall qualitative phase diagram remains unchanged even when the fine-tuned quantities are varied.}).

The QFEM model we used coupled STO's carriers to its soft polar phonon via a dynamical Rashba spin-orbit interaction. We used the simplest model possible, limiting ourselves to just one bosonic mode and one electronic band, describing the ordered state with just a quartic term,  and neglecting any feedback of the ordered state on electronic properties. However, since the Eliashberg equations are both solvable analytically and cheap to compute numerically, it is easy to generalize them to something more realistic for added accuracy and predictive power.  
For example, two straightforward modifications would be to introduce the transverse phonon splitting that occurs in STO due to its tetragonal phase transition, and add additional subleading contributions to pairing, e.g. from the longitudinal mode or  acoustic modes~\cite{fauque2022}. 
Another possibility for further work is to properly describe the evolution of the FE order parameter in metallic STO. This would determine the peak position of the superconducting order parameter in the ordered state, a feature not captured by the simplified quartic theory we employed for the optical phonon in this study.

A major achievement of our theory is to clarify the role that nonlinear processes play in the overall shape of the phase diagram. Specifically, we showed that at the top of the dome fluctuations due to quadratic fermion-boson coupling are parameterically smaller than those from linear coupling, but that they induce corrections in the ordered state. These corrections both move the dome away from the QCP and remove the cusp which  is a universal property of linearly-coupled QC theories. We did so by reinterpreting the quadratic coupling as a correction to the Rashba-like linear channel that provides the pairing attraction in our model.
This type of correction should be universal. Indeed, as a general rule quadratic boson-fermion coupling is in the $A_{1g}$ representation of the lattice symmetry, which means it will contribute attractively in any given pairing channel, no matter its specific symmetry, once the system enters an ordered state. Experimentally, the top of the dome is very rarely ``exactly'' above the putative QCP. Whether it shifts towards or away from the ordered state seems to vary even in subclasses of well known superconductors, e.g. the Fe-based ones and the cuprates (see e.g. Refs. \cite{Taillefer2010,Keimer2015,Fernandes2014,Wang2021,Shiferaw22} and references therein). 
Furthermore, to our knowledge none of the superconductors suspected of being due to QC fluctuations exhibit a cusp at the top of the dome. Our treatment of the nonlinear coupling opens pathways to understand this phenomenology.

In our work, we accounted only for the nonlinear changes induced by static order in our specific el-ph symmetry channel. As discussed above, using a more realistic treatment of the phonon contribution can also be extended to the quadratic coupling. One of the major things we did \emph{not} do in this work is to properly account for the quadratic-coupling fluctuations that dominate at the ultra low density limit, where our theory fails in a completely expected manner~\cite{ruhman2016superconductivity}. On the other hand, at these dopings, quadratic coupling does successfully describe the pairing of STO before it is doped towards the FE transition \cite{ngai1974,vandermarel2019,feigelman_STO_1,volkov2022}, raising the possibility that including the fluctuations will fully reconstruct the phase diagram. We leave to further work a detailed analysis of the crossover between linear and nonlinear attraction, since it requires a proper calculation of the properties of the quadratic coupling at the QCP itself. 

Finally, let us briefly mention what other systems we believe can be studied using analogs of our theory. First and foremost, it is completely unnecessary to limit oneself to inversion-breaking fluctuations. QC Eliashberg methods have been extensively applied to both Fe- and Cu- based superconductors, and it should be straightforward to generate phase diagrams for these systems and see how well they agree with experiment. In the specific context of inversion-breaking systems, a variety of 2D ferroelectric metals exist, such as SnTe or MoTe$_2$ ~\cite{Qi2016,Jindal2023,Novak2013,Parfenev2001}, where a similar treatment may be relevant. 
We believe our work opens avenues to make progress on understanding all such systems. We also stress that our theory made falsifiable predictions only about the superconducting $T_c$, and that a complete verification of the central role of the Rashba QFEM requires making additional falsifiable predictions, ones that may be better tested in systems other than STO.

\section{Acknowledgments}
We thank D. Van Der Marel,  Y. Tomioka, I. Inoue, J. Schmalian, S. Stemmer, J. Harter and J. Sous for helpful discussions, and Y. Tomioka and I. Inoue for sharing unpublished data with us.  We acknowledge support by the Israel Science Foundation (ISF), and the Israeli Directorate for Defense Research and Development (DDR\&D) under grant No. 3467/21.  M.N.G is supported by the Ramon y Cajal Grant RYC2021-031639-I funded by MCIN/AEI/ 10.13039/501100011033 and by the European Union NextGenerationEU/PRTR.


\appendix
\onecolumngrid
\section{\label{app1} Pairing in 3D QFEMs for the linear coupling}
In this appendix we review the theoretical tools used to compute the three-dimensional phase diagram. We follow Ref.~\cite{avi_qfem} closely and elaborate in detail when needed. We mainly focus on the disordered side of the FE transition and comment on how we extend the analysis to the ordered side.

\subsection{\label{appA1} Self-energies}
We start with the bosonic and fermionic self-energies, which are computed to one-loop order. The bosonic self-energy is obtained following the one-loop fermionic bubble as shown diagrammatically in Fig.~\ref{fig_bubble_diag}(a). 
Following Ref.~\onlinecite{avi_qfem} we construct a vector basis and assign three orthonormal vectors corresponding to each vector $\bm k$ as $\hat{k}_t$, $\hat{k}_u$, $\hat{k}$, as shown in Fig.~\ref{fig_bubble_diag}(b).
We define $\hat{k}_t$ as $\hat{k}_t=\hat{z}\times \hat{k}/\vert \hat{z}\times \hat{k} \vert$ to ensure it lies in the $xy$ plane. We then choose $\hat{k}_u$ as $\hat{k}_u = \hat{k}_t \times \hat{k}$, such that its projection on the $xy$ plane is parallel to the projection of $\hat{k}$ on the same plane. 
Expressing the form-factor in Eq.~\eqref{eq:e-ph_coulping_lag} in the new coordinate space, we obtain
\begin{equation}
 \begin{aligned}
	 \hat{k} \times \bm{\sigma} \rightarrow \hat{k}_t \left (\sin\theta \sigma_x - \cos\theta\sigma_y \right) - \hat{k}_u \sigma_z\,.
 \label{eq:form_factor_co_trans}
  \end{aligned}
 \end{equation}

\begin{figure}
\centering
\includegraphics[width=0.8\hsize]{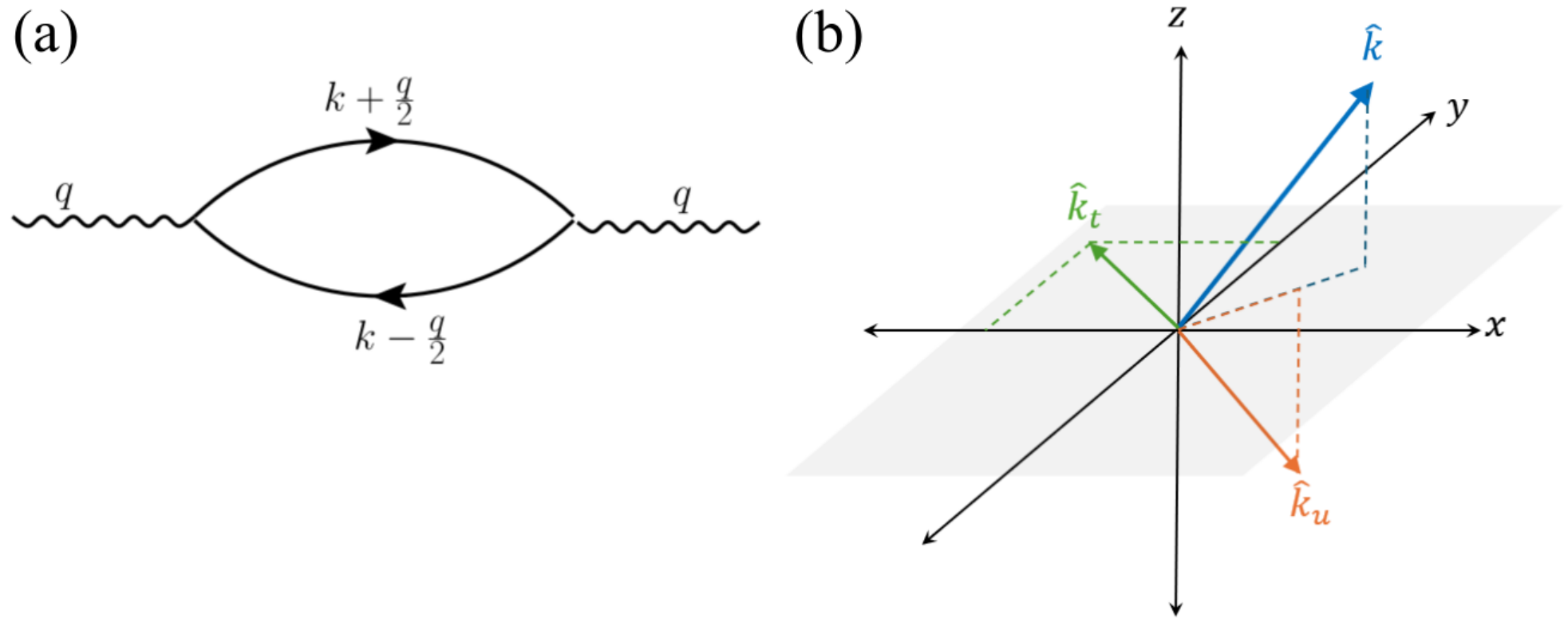}\caption{\label{fig_bubble_diag}(a) The diagram of the one-loop polarization bubble.  The curvy
lines represent the boson propagator and the straight lines with the arrow represent the electronic Green’s function. (b)	The basis for describing the transverse mode. The components transverse to $\hat{k}$ are $\hat{k}_t$ and $\hat{k}_u$. }
\end{figure}
where $\theta$ is the polar angle. The polarization bubble assumes the form 
\begin{equation}
 \begin{aligned}
	 \hat{\Pi}(q) &=\frac{\bar{g}}{k_a^3}T \, \text{Tr}\sum_k \left(\hat{k}_t \left (\sin\theta \sigma_x - \cos\theta\sigma_y \right) - \hat{k}_u \sigma_z\right) \left[ G(k+q/2) G(k-q/2) \right]\left(\hat{k}_t \left (\sin\theta \sigma_x - \cos\theta\sigma_y \right) - \hat{k}_u \sigma_z\right) \\
	 &=\frac{\bar{g}}{k_a^3}T \, \sum_k G(k+q/2) G(k-q/2) \left( \hat{k}_t\hat{k}_t+\hat{k}_u\hat{k}_u\right)\, ,
 \label{eq:bubble_3D_ff}
  \end{aligned}
 \end{equation}
 where $G(k)$ is the fermionic propagator. 
 The fermionic and bosonic four-vectors are denoted by $k=(k_0,\mathbf{k})$ and $q=(q_0,\mathbf{q})$, respectively, with $k_0$ and $q_0$ representing Matsubara frequencies, and $\mathbf{k}$ and $\mathbf{q}$ representing 3D momentum.  
 The trace in the first step in Eq.~\eqref{eq:bubble_3D_ff} is over spin indices. The second step is obtained following spin summations.

 The bubble can then be parameterized as 
 \begin{equation}
 \begin{aligned}
	 \hat{\Pi}(q) =\Pi(q) \hat{P}(\hat{q})\,,
 \label{eq:bubble_3D_trans}
  \end{aligned}
 \end{equation}
where $\hat P(\hat q)$ is the projector on to the plane transverse to $\boldsymbol q$. 
The magnitude of the bubble $\Pi(q) $, is then given by 
\begin{equation}
 \begin{aligned}
\Pi(q) &=\frac{\bar{g}}{k_a^3}T \, \sum_k G(k+q/2) G(k-q/2)  \\
&=\frac{\bar{g}\nu_F}{4\pi} \int \left(\frac{1}{(k_0+q_0/2)+i\epsilon_{\bm{k}+\bm{q}/2}}\right) \left( \frac{1}{(k_0-q_0/2)+i\epsilon_{\bm{k}-\bm{q}/2}} \right)  \,  \frac{dk_0}{2\pi} \,  d\epsilon_k \, d(\cos\theta) \, d\phi    \\
&=\frac{\bar{g}\nu_F}{4\pi}  \int \frac{ \left[\theta(-\epsilon_{\bm{k}})-\theta(-\epsilon_{\bm{k}+\bm{q}})\right]}{ v_F \vert \bm{q} \vert \, \hat{k}\cdot\hat{q}   -i q_0}    \, d\epsilon_k \, d(\cos\theta) \, d\phi    \\
&=\bar{g}\nu_F \left[ 1-  \frac{\vert q_0 \vert}{v_F \vert \bm{q} \vert} \arctan{\left(\frac{v_F \vert \bm{q} \vert }{\vert q_0 \vert }\right)}      \right]\, .
\label{eq:bubble_3D_full}
 \end{aligned}
 \end{equation}
This result gives Eq.~\eqref{eq:bubble_3D_full_mn} of the main text.
Here, we have converted from momentum to energy integration
\begin{equation}
 \begin{aligned}
\frac{\bar{g}}{k_a^3}  \frac{d^3k}{\left(2\pi\right)^3}&=\frac{\bar{g}}{4\pi}    \left(\frac{m^* k_F }{2\pi^2 k_a^3}\right)  \left(\frac{k\, dk}{m^*}\right) d(\cos\theta) \, d\phi \\
&= \frac{\bar{g} \nu_F}{4\pi}  d\epsilon_k \, d(\cos\theta) \, d\phi\,,
 \label{eq:pi_integration_decompose_3D}
  \end{aligned}
 \end{equation}
where $\nu_F=k_F^2/2\pi^2 v_F k_a^3$  is the 3D density of states per spin at the Fermi level and $k_a=1/a$.
In the limit $\vert q_0 \vert \ll  v_F \vert q \vert$ the polarization bubble assumes the form  
\begin{equation}
 \begin{aligned}
\Pi(q) &\approx \bar{g}\nu_F \left( 1- \frac{\pi}{2}  \frac{\vert q_0 \vert}{v_F \vert \bm{q} \vert} \right) \, ,
 \label{eq:bubble_3D_LD}
  \end{aligned}
 \end{equation}
 where we have used the identity $\left[\arctan(x)+\arctan(\frac{1}{x}) \right]=\frac{\pi}{2}$. Eq.~\eqref{eq:bubble_3D_LD} is the Landau damping, which played an important role in the main text discussion in cutting off the effectiveness of the non-linear term in pairing.

Next, we focus on the fermionic self-energy in the normal state. It is shown diagrammatically in Fig.~\ref{self_diag}. It assumes the following form after the spin summation and projection on the transverse sector.
\begin{figure}
\begin{minipage}{0.49\hsize}
    \includegraphics[height=100pt]
    {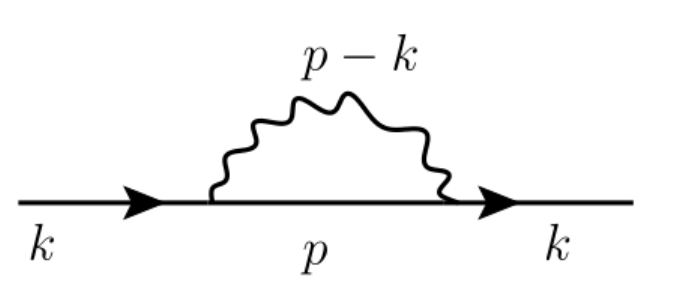}
\caption{\label{self_diag} The diagram of the fermionic self-energy.}
\end{minipage}
\begin{minipage}{0.49\hsize}
\includegraphics[height=100pt]{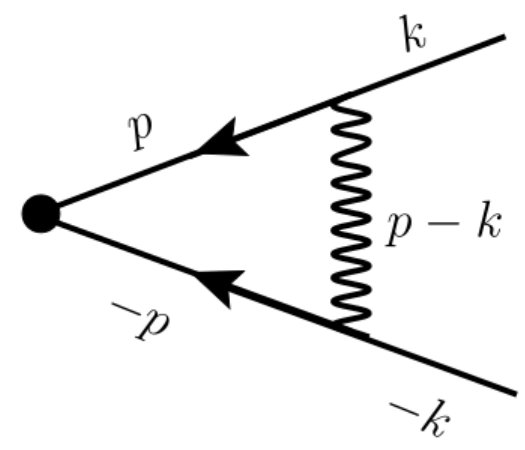}
\caption{\label{vertex_diag} The diagrammatic representation of the pairing vertex.}
\end{minipage}
\end{figure}

 \begin{equation}
 \begin{aligned}
         \Sigma(k_0) &=\frac{\bar{g}}{ k_a^3}  T \sum_{p_0}  \int G(p)D(p-k)  \, \frac{d^3p}{(2\pi)^3} \\
         &=\frac{\bar{g} }{v_F k_a} \frac{1}{(2\pi)^3}  T \sum_{p_0} d(p_0-k_0)  \int d\epsilon \frac{1}{i p_0-\epsilon} \\
         &=\frac{-i\bar{g}\nu_F }{4 k_F^2 a^2}   T \sum_{p_0} d(p_0-k_0)  \text{sgn}(p_0) \, ,
          \label{eq:fermion_self_energy}
 \end{aligned}
\end{equation}
where we have factorized the integration over 3D momentum $d^3p$ to components parallel and transverse to the Fermi surface
 \begin{equation}
 \begin{aligned}
\frac{\bar{g}}{ k_a^3} \frac{d^3p}{(2\pi)^3}=
   \left(\frac{1}{k_a^2}\right)  \frac{1}{4\pi} \frac{\bar{g}\nu_F}{k_F^2 a^2}d^2p_\parallel  d \epsilon_\perp \, .
          \label{eq:gap_integration_decompose_3D}
 \end{aligned}
\end{equation}
Here, the momentum component parallel to the Fermi surface is denoted by $p_\parallel$ and energy is denoted by $\epsilon_\perp=v_F p_\perp$ with $p_\perp$ being the momentum component perpendicular to the Fermi surface.
In Eq.~\eqref{eq:fermion_self_energy}, $D$ represents the bosonic propagator, and the function, $d(\Omega)$, of the bosonic frequency $\Omega=p_0-k_0$ is obtained by integrating the bosonic propagator $D(\Omega,p_\parallel)$ over the momentum $d^2p_\parallel$ up to the Brillouin zone boundary $\Lambda=\pi/a$
\begin{equation}
 \begin{aligned}
        \frac{1}{k_a^2} d(\Omega) =\int_0^{\Lambda/k_a} \frac{p_\parallel dp_\parallel}{r+\vert \bm{p_\parallel} \vert^2+\left(\frac{\Omega}{ck_a}\right)^2 +\frac{\bar{g} \nu_F  \vert \Omega \vert}{ v_F k_a \vert \bm{p_\parallel} \vert}\arctan{\left(\frac{ v_F k_a \vert \bm{p_\parallel} \vert}{ \vert \Omega \vert}\right)}  } \, ,
          \label{eq:boson_function_full}
 \end{aligned}
\end{equation}
where we have used the expression for the polarization bubble in Eq.~\eqref{eq:bubble_3D_full}, renormalize $r$ as $r=r-\bar{g} \nu_F$, and rescale the integration variable as $p_\parallel/k_a \rightarrow p_\parallel$. For simplicity, we neglect the bosonic frequency term $(\Omega/ck_a)^2$ in Eq.~\eqref{eq:boson_function_full}. Then, the bosonic function attains the following form at the QCP ($r=0$) and in the limit $\vert \Omega \vert\ll  v_F \vert p_\parallel \vert$.
  \begin{equation}
 \begin{aligned}
         \frac{1}{k_a^2} d(\Omega)&=\int_0^{\Lambda/k_a} \frac{q dq}{\vert \bm{q} \vert^2+\frac{\pi \bar{ g} \nu_F}{2} \frac{ \vert \Omega \vert}{ v_F k_a \vert \bm{q} \vert} }  = \ln\left[ 1+   \left(\frac{\Lambda}{k_a}\right)^3 \frac{2 v_F k_a}{\pi \bar{ g} \nu_F \vert \Omega \vert} \right]^{1/3} \, .
          \label{eq:boson_function_QCP}
 \end{aligned}
\end{equation}
Inserting this into Eq.~\eqref{eq:fermion_self_energy}, we obtain expression for the fermionic self energy at the QCP
\begin{equation}
 \begin{aligned}
         \Sigma(k_0) &=-i\frac{\bar{g} }{4\pi^2 v_F k_a}  k_0 \ln \left(\frac{\omega_\Lambda}{ \vert k_0 \vert }\right)^{1/3}\,,
          \label{eq:fermion_self_energy_QCP}
 \end{aligned}
\end{equation}
where we have defined $\omega_\Lambda=2 v_F \Lambda^3/\pi \bar{ g} \nu_F k_a^2$. This result gives Eq.~\eqref{eq:fermion_self_energy_QCP_mn} of the main text.

\subsection{\label{appA2} Gap equation}
In this section, we focus on the linearized equation for the pairing vertex, shown diagrammatically in Fig.~\ref{vertex_diag}. It assumes the form
\begin{equation}
 \begin{aligned}
	 \Phi_{\alpha \beta}(k) =\frac{\bar{g} T}{ k_a^3}    \sum_{p_0}\int & \frac{d^3p}{(2\pi)^3}  G(p)G(-p)D(p-k)    \,  \bm{\gamma}^{ \nu \alpha}(-\hat{k})\Phi_{\nu \mu}(k)\cdot  \hat{P}(\hat{q}) \cdot \bm{\gamma}^{ \mu \beta}(\hat{k}) \, ,
 \label{eq:gapeq_vector1}
 \end{aligned}
\end{equation}
where the interaction form factor, arising due to the vector coupling, is represented by $\bm{\gamma}(\hat{k})=\hat{k} \times \bm{\sigma}$. It can be decomposed in the vector basis $\hat{k}_t$, $\hat{k}_u$, $\hat{k}$ (discussed in Appendix~\ref{appA1}) as $\gamma(\hat{k})=\hat{k} \times \bm{\sigma}=\sigma_u \hat{k}_t - \sigma_t \hat{k}_u$, where $\sigma_t=\hat{k}_t \cdot \bm{\sigma}$ and $\sigma_u=\hat{k}_u \cdot \bm{\sigma}$.
Since $\hat{q}$ is approximately parallel to the FS, lying in a plane spanned by $\hat{k}_u$ and $\hat{k}_t$, we can express  $\hat{q}$ as $\hat{q} \approx \cos \phi \hat{k}_u+ \sin \phi \hat{k}_t$. Additionally, the gap function, $ \Phi(k)$, can be decomposed as a sum over irreducible representations $ \Phi(k)=i \sigma_y \sum_{nj} \phi_{nj}(k_0)F_n^j(\hat{k})$ ~\cite{avi_qfem}. The leading attractive pairing channel in these systems is well known to be singlet ~\cite{avi_qfem, maria1}. For further analysis, we consider only the singlet pairing where $n=0$ and $F_n^j=1$. We denote $\phi_{nj}$ at $n=0$ as $\phi$. The spin summations and projection on the transverse sector follows
\begin{equation}
\begin{aligned}
&    \left( -\hat{k} \times \bm{\sigma}  \right)_{ \nu \alpha} \Phi_{\nu \mu}(p_0,\bm{k}) \cdot\hat{P}(q) \cdot  \left( \hat{k} \times \bm{\sigma}  \right)_{ \mu \beta} \\
 & =- \left(   \cos \phi \sigma^T_t- \sin \phi \sigma^T_u   \right)  i \sigma_y  \phi(p_0)  \left(   \cos \phi \sigma_t- \sin \phi \sigma_u   \right)\\
 &= i \sigma_y \phi(p_0) \, ,
 \label{eq:vertex_vector1}
 \end{aligned}
\end{equation}
where we have used the identities $\sigma_j^T i \sigma_y = -i \sigma_y \sigma_j$ and  $(\bm{\sigma}\cdot \bm{a}) (\bm{\sigma} \cdot \bm{b})=(\bm{a}\cdot \bm{b})\sigma_0+(\bm{a} \times \bm{b}) \bm{\sigma}$. Hence we simplify Eq.~\eqref{eq:gapeq_vector1} as
\begin{equation}
 \begin{aligned}
	 \phi(k_0) &=\frac{\bar{g} }{v_F k_a} \frac{1}{(2\pi)^3} T \sum_{p_0} d(p_0-k_0)\phi(p_0) \int_{-\infty}^{\infty} d\epsilon \frac{1}{i\tilde{\Sigma}(p_0)-\epsilon}\, \frac{1}{-i\tilde{\Sigma}(p_0)-\epsilon}\\
	  &=  \frac{\bar{g}\nu_F k_a^2}{4k_F^2} T \sum_{p_0} \frac{ d(p_0-k_0)\phi(p_0)}{\vert \tilde\Sigma(p_0)\vert}\, ,
 \label{eq:vertex}
 \end{aligned}
\end{equation} 
where we have defined $\tilde{\Sigma}(p_0)=p_0+\Sigma(p_0)$.
This result is also shown in Eq.~\eqref{eq:gapeq_vector2_mn} of the main text where the effective coupling constant is defined as $\lambda=\bar{g}\nu_F/4(k_Fa)^2$.

\subsection{\label{appA3} Exclusion of thermal contribution from the gap equation}
The expressions for the self-energy, $\Sigma(k_0)$, and the pairing vertex, $\phi(k_0)$, are presented in Eq.~\eqref{eq:fermion_self_energy} and Eq.~\eqref{eq:vertex}, respectively. Both expressions contain terms with $k_0=p_0$ that contribute divergently at the QCP. In this section, we discuss the procedure for discarding this divergent term.

Within the framework of Eliashberg theory, the fermionic full Green’s function, $\hat{G}$,  and self-energy, $\hat{\Sigma}$, in Nambu space can be expressed as ~\cite{moon2010}
\begin{equation}
 \begin{aligned}
	&\hat{\Sigma}(k_0)=-i\Sigma(k_0) \hat{\tau}_0-\phi(k_0)\hat{\tau}_1 \, ,\\
	&\hat{G}^{-1}(\epsilon_k,k_0)=\hat{G}_0^{-1}(\epsilon_k,k_0)-\hat{\Sigma}(k_0)=i\tilde\Sigma(k_0)\hat{\tau}_0-\epsilon_k\hat{\tau}_3+\phi(k_0)\hat{\tau}_1 \, .
	 \label{eq:green_nambu}
 \end{aligned}
\end{equation}
Here, $\hat{\tau}$ are the Pauli matrices in the particle-hole space. $\epsilon_k$ represents the fermionic dispersion in the normal state. $\hat{G}_0$ is the Green's function for the non-interacting fermions. $\phi(k_0)$ is the pairing vertex and $\Sigma$ is the regular self-energy (not in Nambu space).
 $\hat{\Sigma}$ can be written using $\hat{G}$  as
\begin{equation}
 \begin{aligned}
	\hat{\Sigma}(k_0)&=\frac{\lambda}{\pi}  T \sum_{p_0} d(p_0-k_0) \int d\epsilon_p \hat{G}(p_0,\epsilon_p)\, , \\
	-i\Sigma(k_0) \hat{\tau}_0-\phi(k_0)\hat{\tau}_1    &= \frac{\lambda}{\pi}  T \sum_{p_0} d(p_0-k_0) \int d\epsilon_p \frac{-i\tilde{\Sigma}(p_0)\hat{\tau}_0-\epsilon_p\hat{\tau}_3+\phi(p_0)\hat{\tau}_1 }{\tilde{\Sigma}^2(p_0)+\epsilon_p^2+\phi^2(p_0)} \, .
	 \label{eq:self_nambu}
 \end{aligned}
\end{equation}
where $\lambda$ was defined in \eqref{eq:param_lambda}. 
We have also decomposed the momentum into components parallel and perpendicular to the Fermi surface. as discussed in Appendix~\ref{appA1}.
Comparing the coefficients of the Pauli matrices in Eq.~\eqref{eq:self_nambu}, we obtain the following set of Eliashberg equations.
\begin{equation}
 \begin{aligned}
	\Sigma(k_0)    &= \frac{\lambda}{\pi}  T \sum_{p_0} d(p_0-k_0) \tilde{\Sigma}(p_0) \int d\epsilon_p \frac{1 }{\tilde{\Sigma}^2(p_0)+\epsilon_p^2+\phi^2(p_0)}\\
	&=\frac{\lambda}{\pi} T  \sum_{p_0} d(p_0-k_0) \tilde{\Sigma}(p_0) \frac{\pi}{\sqrt{\tilde{\Sigma}^2(p_0)+\phi^2(p_0)}}\, ,\\
	\phi(k_0)&=\frac{\lambda}{\pi} T  \sum_{p_0} d(p_0-k_0) \phi(p_0) \frac{\pi}{\sqrt{\tilde{\Sigma}^2(p_0)+\phi^2(p_0)}}\, .
	 \label{eq:Eliashberg_der}
 \end{aligned}
\end{equation}
The coupled Eliashberg equations for self-energy and pairing vertex is then given by ~\cite{moon2010, chubukov2020}
\begin{equation}
 \begin{aligned}
	\tilde{\Sigma}(k_0)=k_0+\lambda T\sum_{p_0}d(p_0-k_0)\frac{\tilde{\Sigma}(p_0)}{\sqrt{ \tilde{\Sigma}^2(p_0)+ \phi^2(p_0)   }} \, .
	 \label{eq:Eliashberg_self}
 \end{aligned}
\end{equation}

\begin{equation}
 \begin{aligned}
	\phi(k_0)=\lambda T\sum_{p_0}d(p_0-k_0)\frac{\phi(p_0)}{\sqrt{ \tilde{\Sigma}^2(p_0)+ \phi^2(p_0)   }}\, .
	 \label{eq:Eliashberg_vertex}
 \end{aligned}
\end{equation}

The interaction term $d(p_0-k_0)$ has a diverging contribution at $k_0=p_0$ at the QCP (see Eq.~\eqref{eq:boson_function_QCP}), which represents thermal fluctuation~\cite{chubukov2005,avi_qfem}. This is similar to the effect of non-magnetic impurities and should not affect $T_c$ according to Anderson's theorem ~\cite{anderson1959}. We remove the thermal contribution in the gap equation and self-energy by subtracting the term with $k_0=p_0$ from Eq.~\eqref{eq:Eliashberg_self} and Eq.~\eqref{eq:Eliashberg_vertex} in the following manner ~\cite{sachdev1988,chubukov2008,moon2010, chubukov2020} :
\begin{equation}
 \begin{aligned}
	 &\tilde{\Sigma}^\prime(k_0)= \tilde{\Sigma}(k_0)(1-Q(k_0))\, , \\
	 &\phi^\prime(k_0)=\phi(k_0) (1-Q(k_0))\, , 
	 \label{eq:thermal}
 \end{aligned}
\end{equation}
where,
\begin{equation}
 \begin{aligned}
 Q(k_0)=\lambda T\frac{\,d(0)}{\sqrt{ \tilde{\Sigma}^2(k_0)+ \phi^2(k_0) }} \, .
  \label{eq:thermal1}
 \end{aligned}
\end{equation}
 At this stage, a new gap function $\Delta(k_0)=k_0 \, \phi(k_0)/\tilde{\Sigma}(k_0)$ is convenient~\cite{sachdev1988,chubukov2008,moon2010, chubukov2020} since it is invariant under the transformation in Eq.~\eqref{eq:thermal} ($\phi(k_0)/\tilde{\Sigma}(k_0)=\phi^\prime(k_0)/\tilde{\Sigma}^\prime(k_0)$). Thus, after excluding of the thermal contribution and incorporating the gap function, $\Delta(k_0)$, Eq.~\eqref{eq:Eliashberg_self} takes the form
\begin{equation}
 \begin{aligned}
	&\tilde{\Sigma}^\prime(k_0)=k_0+\lambda T\sum_{p_0\ne k_0}d(p_0-k_0)\frac{p_0}{\sqrt{ p_0^2+ \Delta^2(p_0) }}\, ,\\
	& \frac{\tilde{\Sigma}^\prime(k_0)\Delta(k_0)}{k_0}= \phi^\prime(k_0)=\Delta(k_0)+\lambda T \sum_{p_0 \ne k_0}d(p_0-k_0)\frac{\Delta(k_0) p_0/k_0}{\sqrt{ p_0^2+ \Delta^2(p_0) }}\, .
	 \label{eq:Eliashberg_self_del}
 \end{aligned}
\end{equation}
Similarly, Eq.~\eqref{eq:Eliashberg_vertex} takes the form
\begin{equation}
 \begin{aligned}
	\phi^\prime(k_0)=\lambda T\sum_{p_0 \ne k_0}d(p_0-k_0)\frac{\Delta(p_0)}{\sqrt{ p_0^2+ \Delta^2(p_0)   }}\, .
	 \label{eq:Eliashberg_vertex_del}
 \end{aligned}
\end{equation}
Finally,  Eq.~\eqref{eq:Eliashberg_self_del} and Eq.~\eqref{eq:Eliashberg_vertex_del} lead to
\begin{equation}
 \begin{aligned}
	\Delta(k_0)&=\lambda T\sum_{p_0\ne k_0} d(p_0-k_0)   \frac{\Delta(p_0)-\Delta(k_0) \frac{p_0}{k_0}}{\sqrt{ p_0^2+ \Delta^2(p_0) }} \\
	&=\lambda T \sum_{p_0>0, p_0\ne k_0} \left[ \frac{\Delta(p_0)}{\sqrt{ p_0^2+ \Delta^2(p_0) }}\left( d(p_0-k_0) + d(p_0+k_0)    \right) - \frac{\Delta(k_0)}{k_0\sqrt{ p_0^2+ \Delta^2(p_0) }}\left(p_0 d(p_0-k_0) -p_0 d(p_0+k_0) \right) \right] \, .
	 \label{eq:Eliashberg_vertex_thermal}
 \end{aligned}
\end{equation}
Thus two self-consistent equations (Eq.~\eqref{eq:Eliashberg_self} and Eq.~\eqref{eq:Eliashberg_vertex}) are reduced to a single expression (Eq.~\eqref{eq:Eliashberg_vertex_thermal}).
The linearized gap equation can be expressed as
\begin{equation}
 \begin{aligned}
	\Delta(k_0)&=  \lambda T\sum_{p_0\ne k_0} d(p_0-k_0)  \left(  \frac{\Delta(p_0)}{\vert p_0 \vert}  -\frac{\Delta(k_0)}{k_0} \text{sgn}(p_0) \right)   \\
	&=\lambda T \sum_{p_0>0, p_0\ne k_0}\left[ \frac{\Delta(p_0)}{p_0} \left( d(p_0-k_0)+  d(p_0+k_0)  \right) - \frac{\Delta(k_0)}{k_0} \left( d(p_0-k_0)- d(p_0+k_0) \right) \right]\, .
	 \label{eq:Eliashberg_vertex_thermal_linear}
 \end{aligned}
\end{equation}
$\Delta(k_0)$ in Eq.~\eqref{eq:Eliashberg_vertex_thermal} and Eq.~\eqref{eq:Eliashberg_vertex_thermal_linear} appears on both left and right hand sides of the equation.  Eq.~\eqref{eq:Eliashberg_vertex_thermal_linear} resembles an eigenvalue equation: 
\begin{equation}
 \begin{aligned}
	\Delta(k_0)=\frac{\lambda T \sum_{p_0 \ne k_0} \frac{1}{p_0} \left[ d(p_0-k_0)+  d(p_0+k_0)  \right]\Delta(p_0)}{1+\lambda T\frac{1}{k_0}\sum_{q_0 \ne k_0}  \left[ d(q_0-k_0)-  d(q_0+k_0)  \right] } \, .
	 \label{eq:eliashberg_gapeq_matrix}
 \end{aligned}
\end{equation}
Here, the upper cutoff in the frequency summation is the Fermi energy $E_F/k_B$. The values of the parameters going into Eq.~\eqref{eq:eliashberg_gapeq_matrix} are discussed in Section~\ref{sec2c}.
We solve this equation numerically to calculate $T_c$ and the results are discussed in Section~\ref{sec3} of the main text.

\section{\label{app2} Dependence of the linear Rashba Coupling, $g_{TO}$, on density, $k_Fa$}

The linear coupling, $g_{TO}$, discussed in Section~\ref{sec2c}, was determined in Ref.~\cite{maria2} through \emph{ab initio} calculations of the band splitting in STO under a polar distortion. These calculations reveal a dome-like dependence of the coupling on density, $k_Fa$, as described by the following expression ~\cite{maria2}.
\begin{equation}
 \begin{aligned}
	g_{TO}=\frac{\Gamma \sin{\frac{k_Fa}{\sqrt{2}}}}{\sqrt{ \sin^4{\frac{k_Fa}{\sqrt{2}}}+\beta^2 }} \, ,
	 \label{eq:param_gto}
 \end{aligned}
\end{equation}
where the parameters take the values $\Gamma \approx 20.24$ meV and $\beta \approx 0.09$. This relationship is also illustrated in Fig.~\ref{fig_gto_DFT}, which shows the coupling constant, $g_{TO}$, as a function of carrier density, $n_e=k_F^3/3\pi^2$.

\section{\label{app3} Nonlinear coupling}
The QC Eliashberg theory predicts that the linear coupling favors the maximum $T_c$ at the QCP where it exhibits a cusp (except for small values of $\omega_0^2$). However, this behavior contradicts experimental observations in Ba/Ca-doped STO, where the peak of the dome lies within the ordered phase ~\cite{tomioka2022}. Therefore, linear coupling alone cannot fully account for SC in Ba/Ca-doped STO. In this appendix, we explore beyond the linear order, i.e., the coupling to a pair of TO modes, which introduces a correction to the linear coupling, thereby enhancing the same.
The main results are presented in Section~\ref{sec4}, while the corresponding technical details are discussed here.

\subsection{\label{app3a} The relative contribution of the linear and nonlinear coupling to pairing}
In this appendix, we estimate the pairing fluctuation associated with the two phonon mechanism (described by the Lagrangian in Eq.~\eqref{eq:e-ph_vec_quad_coulping_lag}) and compare it with the same associated with the linear coupling near the top of the superconducting dome, highlighting their relative contribution to driving superconductivity in the QFEM model.
\begin{figure}
\includegraphics[width=0.3\hsize]{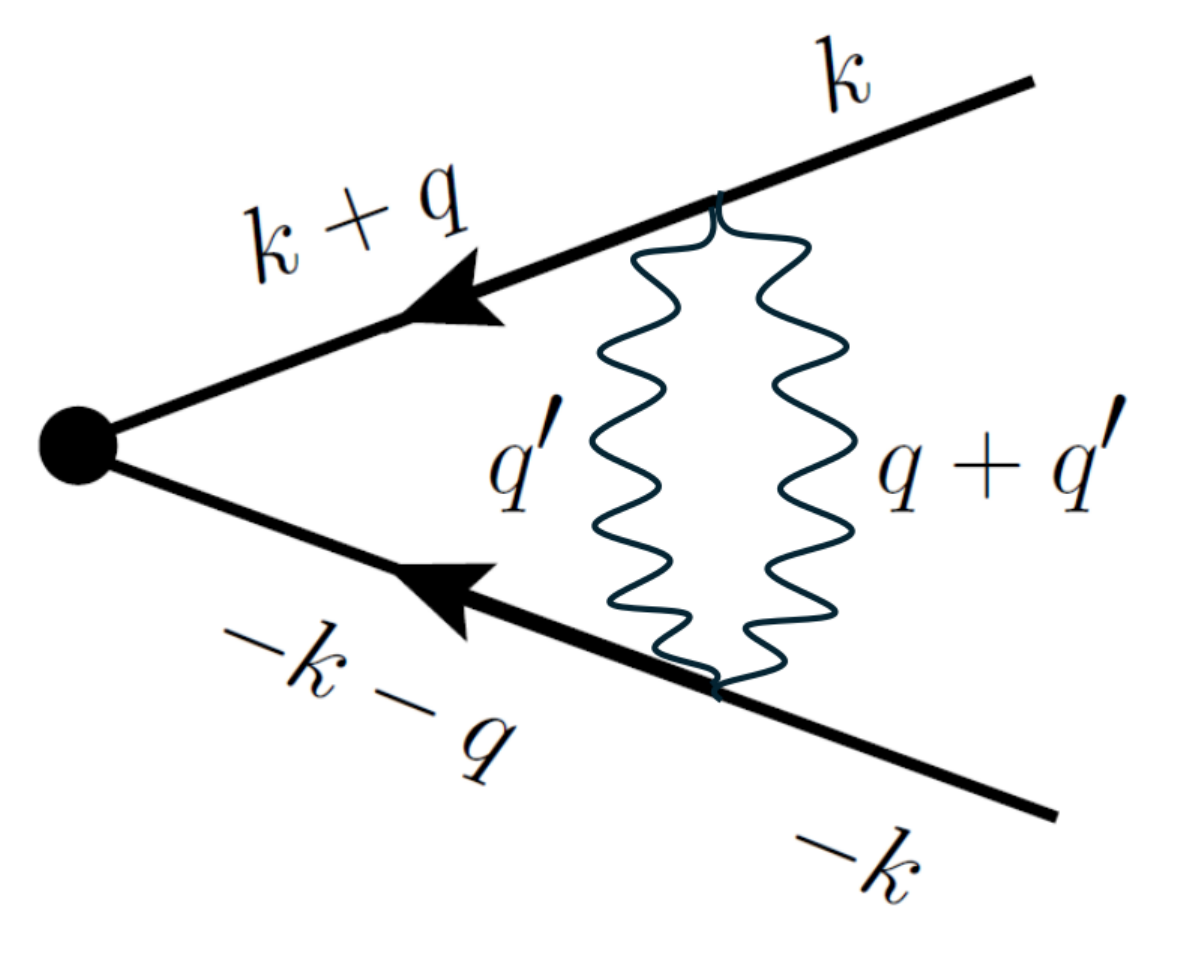}
\caption{\label{vertex_diag_{NL}} The diagrammatic representation of the pairing vertex for nonlinear coupling.}
\end{figure}

The vertex diagram for the quadratic coupling is shown in Fig.~\ref{vertex_diag_{NL}}. The gap equation at the QCP assumes the form
\begin{equation}
 \begin{aligned}
\phi(k_0) &=\frac{\lambda^\prime v_F}{\pi} \int dq_0 \int d^3q  \,G(k+q)G(-k-q)\phi(k_0+q_0) \frac{1}{(2\pi)^4 k_a}   \int dq_0^\prime \int d^3q^\prime \,  D(q^\prime)D(q+q^\prime)\\
 &=\lambda^\prime \int dq_0   \frac{\phi(k_0+q_0) }{\vert \tilde\Sigma(k_0+q_0)\vert} \int d^2q \,  \frac{1}{(2\pi)^4 k_a}  \int dq_0^\prime   \int d^3q^\prime \,  D(q^\prime)D(q+q^\prime)\\
 &=\lambda^\prime \int dq_0   \frac{\phi(k_0+q_0) }{\vert \tilde\Sigma(k_0+q_0)\vert} d_{NL}(q_0) \, .
	 \label{eq:gapeq_quad1}
 \end{aligned}
\end{equation}
Here, we denote the quadratic coupling constant as $\lambda^\prime$, without delving into additional details in this appendix. For the detailed discussion of the coupling constant, please see Section~\ref{sec4} (where we demonstrate $\lambda^\prime=\lambda_{NL}  \langle \bm{\eta} \rangle^2$).
The momentum integration in Eq.~\eqref{eq:gapeq_quad1} is factorized as
$d^3q \rightarrow (1/v_F)d^2q_\parallel dq_\epsilon$, where $q_\parallel$ is parallel to the Fermi surface and $q_\epsilon$ is along the energy axis. We also drop the index ($\parallel$) in $q_\parallel$, denoting it as $q$. Furthermore, the momenta are rescaled as $\bm{q}/k_a \rightarrow \bm{q}$ and  $\bm{q}^\prime/k_a \rightarrow \bm{q}^\prime$.
In the last step of Eq.~\eqref{eq:gapeq_quad1}, we define a frequency dependent kernel, $d_{NL}(q_0)$, which is given by
\begin{equation}
 \begin{aligned}
 d_{NL}(q_0)=\frac{1}{(2\pi)^4 k_a} \int dq_0^\prime  \int d^2q \int d^3q^\prime \,  D(q^\prime)D(q+q^\prime)\, .
 \label{eq:gapeq_boson_quad1}
 \end{aligned}
\end{equation}
Here, the bosonic propagators, $D(q^\prime)$ and $D(q+q^\prime)$, have a contribution from the Landau damped polarization bubble ($\delta\Pi(q^\prime)$ and $\delta\Pi(q+q^\prime)$, respectively).

To estimate the relative strength of pairing fluctuation from linear and quadratic couplings, we calculate the magnitude of the function $d_{NL}(q_0)$ at the top of the dome and compare it with the bosonic function, $d(q_0)$, associated with the linear coupling.
The magnitude of $d_{NL}(q_0)$ is then given by
\begin{equation}
 \begin{aligned}
& d_{NL}(q_0)=  \frac{1}{(2\pi)^4 k_a v_F} \int dq_0^\prime   \int d^2q^\prime \int d^2q   \int_{-\infty}^{\infty} dq^\prime_\epsilon \left( \frac{1}{\vert \bm{q}^\prime \vert^2+{q^\prime_\epsilon}^2+\left( \frac{q_0^\prime \zeta }{c}  \right)^2+\delta\Pi(q^\prime)} \right)  \left( \frac{1}{\vert \bm{q} \vert^2+{q^\prime_\epsilon}^2+\left( \frac{(q_0+q_0^\prime)\zeta}{c}  \right)^2+\delta\Pi(q+q^\prime)}\right)\\
 &=\qquad  \frac{\pi}{(2\pi)^4 k_a v_F}   \int dq_0^\prime    \int d^2q^\prime \int d^2q \\
 & \frac{1}{\left(\vert \bm{q}^\prime \vert^2+\left( \frac{q_0^\prime \zeta }{c}  \right)^2+\delta\Pi(q^\prime) \right)\sqrt{ \vert \bm{q} \vert^2+\left( \frac{(q_0+q_0^\prime)\zeta}{c}  \right)^2+\delta\Pi(q+q^\prime)} +\left( \vert \bm{q} \vert^2+\left( \frac{(q_0+q_0^\prime)\zeta}{c}  \right)^2+\delta\Pi(q+q^\prime)\right)\sqrt{ \vert \bm{q}^\prime \vert^2+\left( \frac{q_0^\prime \zeta }{c}  \right)^2+\delta\Pi(q^\prime)}}\\
 &= \frac{\pi}{(2\pi)^4 k_a v_F}   \int dq_0^\prime    \int d^2q^\prime \frac{\ln\left( \frac{\sqrt{ \vert \bm{q}^\prime \vert^2+\left( \frac{q_0^\prime \zeta }{c}  \right)^2+\delta\Pi(q^\prime) }+\sqrt{ \left( \frac{(q_0+q_0^\prime)\zeta}{c}  \right)^2+\delta\Pi(q+q^\prime) +\left(\frac{\Lambda}{k_a}\right)^2 } }{\sqrt{ \vert \bm{q}^\prime \vert^2+\left( \frac{q_0^\prime \zeta }{c}  \right)^2+\delta\Pi(q^\prime) }+\sqrt{ \left( \frac{(q_0+q_0^\prime)\zeta}{c}  \right)^2+\delta\Pi(q+q^\prime) }}  \right)}{\sqrt{ \vert \bm{q}^\prime \vert^2+\left( \frac{q_0^\prime \zeta }{c}  \right)^2+\delta\Pi(q^\prime)  }} \\
 &=  \frac{\pi}{(2\pi)^4 k_a v_F}  \int dq_0^\prime  F\left( \left(\frac{(q_0+q_0^\prime)\zeta}{c}  \right)^2+\delta\Pi(q+q^\prime), \left( \frac{q_0^\prime \zeta }{c}  \right)^2+\delta\Pi(q^\prime),\frac{\Lambda}{k_a} \right) \, .
 \label{eq:gapeq_boson_quad3}
 \end{aligned}
\end{equation}
Here, the UV cutoff for the integration over momentum is the Brillouin zone boundary $\Lambda=\pi/a$. The component $q^\prime_\epsilon$ is along the energy axis. We have also defined $\zeta=k_B/\hbar k_a$. The function $F(x,y,z)$ has the form
\begin{equation}
 \begin{aligned}
F(x,y,z)&=\left( \sqrt{x}-\sqrt{x+z^2} \right)\ln\left( \sqrt{x}+\sqrt{y} \right) -\left( \sqrt{y}+\sqrt{x+z^2} \right) \ln\left(\frac{\sqrt{y}+\sqrt{x+z^2} }{\sqrt{x}+\sqrt{y} }\right) \\
&+\left(\sqrt{x+z^2}-\sqrt{x} \right) \ln \left( \sqrt{x}+\sqrt{y+z^2} \right)+   \left( \sqrt{x+z^2}+\sqrt{y+z^2} \right) \ln\left(\frac{\sqrt{x+z^2}+\sqrt{y+z^2} }{\sqrt{x}+\sqrt{y+z^2} }\right) \, .
 \label{eq:quad_function}
  \end{aligned}
\end{equation}
At low temperatures, the first two arguments to $F$ in Eq. \eqref{eq:gapeq_boson_quad3} are always much smaller than the third and we can approximate it as,
\begin{equation}
    F(x,y,z) \approx 2\ln(2)z = \mbox{const}.
\end{equation}

Near the top of the SC dome (at $n_e=1\times 10^{20}$ cm$^{-3}$), the Landau damped polarization bubble takes the value $\delta\Pi=(\pi \bar{ g} \nu_F /2) (c/v_F) \approx 0.15$. The Fermi energy at this density is $E_F=18.5$ meV, and $1/v_Fk_a=2 \nu_F \pi^2/(k_Fa)^2$ with $\nu_F=0.24$ eV$^{-1}$. We consider $q_0=2 \pi T_c$ with $T_c \approx 1$ K near the top of the dome. The lower limit of the integration in Eq.~\eqref{eq:gapeq_boson_quad3} is set to $2 \pi T_c$ and the upper limit is the Fermi level, $E_F/k_B$.
We perform the integration numerically, which yields $d_{NL}=4.9 \times 10^{-3}$.

Now, let's focus again on the kernel, $d(q_0)$, associated with the linear coupling (also discussed in Appendix~\ref{appA2}). It can be rewritten as  
\begin{equation}
 \begin{aligned}
	d(q_0)&=\int_0^{\Lambda/k_a} d^2q \left( \frac{1}{\vert \bm{q}\vert^2 +\left( \frac{q_0 \zeta }{c}  \right)^2 +\delta\Pi(q)   }\right)  =\frac{1}{2}\ln \left( 1+\frac{\left(\frac{\Lambda}{k_a} \right)^2}{\left( \frac{q_0 \zeta }{c}  \right)^2 +\delta\Pi(q) } \right) \, .
 \label{eq:gapeq_boson_vector2}
 \end{aligned}
\end{equation}
 The function, $d(q_0)$, is also shown in Eq.~\eqref{eq:boson_function_QCP}, where we neglected the frequency term, $q_0^2\zeta^2/c^2$, from the bosonic propagator. At the top of the dome, Eq.~\eqref{eq:gapeq_boson_vector2} yields a value $d=2.1$. Thus, the ratio of the bosonic kernels associated with the quadratic and linear couplings at the top of the dome is obtained as $d_{NL}/d \sim 2.3 \times 10^{-3}$ which reveals the dominance of the linear coupling over the nonlinear coupling.

\subsection{\label{app3b} The role of Landau damping in the nonlinear coupling at the high density limit}
Based on the results discussed in Section~\ref{sec3} and Section~\ref{sec4}, we conclude that in the moderate to high density regime, the linear coupling plays a dominant role in driving superconductivity in Ba/Ca doped STO. 
Once the FE moment develops a finite expectation value, an additional correction from nonlinear coupling emerges, capturing finer features near the QCP.
However, the non-linear coupling can also contribute to pairing from the fluctuations around the mean (the so called two-phonon coupling or Ngai mechanism).
 Van der Marel et al.~\cite{vandermarel2019} were the first to apply the two-phonon mechanism to the soft FE mode in STO. Through optical conductivity measurements, they estimated the coupling and concluded that it is a relevant mechanism. Kiselov and Feigelman~\cite{feigelman_STO_1} further applied this two-phonon coupling model to the extremely low-density regime ($n_e < 1.5 \times 10^{18}$ cm$^{-3}$), where only one electronic band is occupied. In their treatment of the low-density limit, they made two key approximations: the renormalization of the phonon frequency $\omega_{T}$ due to carrier concentration was neglected, and the effective interaction was assumed to be static. Despite these simplifications, the results were in good agreement with the measured critical temperature $T_c$~\cite{behnia2014}. Volkov et al.~\cite{volkov2022} studied the two phonon coupling close to the quantum-critical point to describe superconductivity in doped STO. However, the results are best justified at the extreme low-density limit, where the Fermi surface is sufficiently small to satisfy the antiadiabetic limit (Fermi energy, $E_F<\omega_T$),  in accord with Ref.~\onlinecite{feigelman_STO_1}. 

In this section we estimate how important is the Landau damping correction in the two-phonon process near the top of the dome. We analyze $d_{NL}(q_0)$ (defined in  Eq.~\eqref{eq:gapeq_boson_quad1}). We perform the integral in a slightly more transparent but approximate fashion by integrating over momenta independently. After the integration over momenta, $q$ and $q^\prime$, up to the Brillouin zone boundary $\Lambda=\pi/a$, the kernel takes the form
\begin{equation}
 \begin{aligned}
  d_{NL}(q_0) &\approx \frac{1}{(2\pi)^4 k_a v_F}  \int dq_0^\prime  \int dq^\prime_\epsilon  \int d^2q^\prime  \left( \frac{1}{\vert \bm{q}^\prime \vert^2+{q^\prime_\epsilon}^2+\left( \frac{q_0^\prime \zeta }{c}  \right)^2+\delta\Pi(q^\prime)} \right)  \int d^2q\left( \frac{1}{\vert \bm{q} \vert^2+{q^\prime_\epsilon}^2+\left( \frac{(q_0+q_0^\prime)\zeta}{c}  \right)^2+\delta\Pi(q+q^\prime)}\right) \\
=& \frac{1}{(2\pi)^4 k_a v_F} \int dq_0^\prime  \int dq^\prime_\epsilon \ln \left[ \frac{\Lambda/k_a}{\left( {q^\prime_\epsilon}^2+\left( \frac{q_0^\prime \zeta}{c}  \right)^2+ \delta\Pi(q^\prime) \right)^{1/2}} \right]   \ln \left[ \frac{\Lambda/k_a}{\left( {q^\prime_\epsilon}^2+\left( \frac{(q_0+q_0^\prime)\zeta}{c}  \right)^2+\delta\Pi(q+q^\prime) \right)^{1/2}} \right]\, .
 \label{eq:gapeq_boson_quad2}
 \end{aligned}
\end{equation}
Here, we have shifted the integration variable $\bm{q}\rightarrow \bm{q}-\bm{q}^\prime$. The UV momentum cutoff is the Brillouin zone boundary $\Lambda=\pi/a$.
The Landau damped polarization bubble ($\delta\Pi(q^\prime)$ and $\delta\Pi(q+q^\prime)$) sets an IR cutoff scale for the momentum integration. 
The results of Ref.~\cite{volkov2022} are reproduced in the limit $v_F \ll c$, and when neglecting the Landau damping. Indeed, in this case Eq.~\eqref{eq:gapeq_boson_quad2} simplifies to $d_{NL}\sim k_F^2 \ln \left({\Lambda}/{ck_F}\right)$.

To extend the solution of $T_c$ to the higher density, it is essential to estimate the typical magnitude of $\delta \Pi$ and determine when it is a relevant cutoff. In the space of frequency and momentum, it dominates the bosonic function $I_{NL}$ when
\begin{equation}
 \begin{aligned}
 \left(\frac{q_0^\prime}{c}\right)^2 <\gamma \frac{\vert q_0^\prime \vert}{\vert \bm{q}^\prime \vert}; \qquad \vert \bm{q}^\prime \vert^2 \sim \gamma \frac{\vert q_0^\prime \vert}{\vert \bm{q}^\prime \vert}.
 	 \label{eq:vcc_an_LD_condition1}
 \end{aligned}
\end{equation}
We recall that $\gamma=\pi \bar{g}\nu_F/2 v_F$ with $\bar{g}=4 {g}_{TO}^2 (k_Fa)^2 D_0$ (see Section~\ref{sec2c}).   
Using these expressions with the assumption that $\vert q_0^\prime \vert/\vert \bm{q}^\prime \sim c$ Eq.~\eqref{eq:vcc_an_LD_condition1} assumes the form 
\begin{equation}
\begin{aligned}
 q_0^\prime < E_F \left(\frac{g_{TO}^2}{\omega_{TO}E_F}  \right)^{1/2} \left( \frac{c}{v_F}  \right)^{3/2} \left(k_Fa\right)^{3/2}.
\label{eq:vcc_an_LD_condition2}
\end{aligned}
\end{equation}
and $\delta \Pi$ takes the value
 \begin{equation}
 \begin{aligned}
\frac{\sqrt{\delta \Pi}}{k_F}  \sim \frac{g_{TO} \sqrt{D_0}}{\sqrt{ \pi E_F} } \left(k_Fa\right)^{3/2} \, .
 	 \label{eq:vcc_an_LD_cutoff1}
 \end{aligned}
\end{equation}
We use Eq.~\eqref{eq:vcc_an_LD_cutoff1} to estimate $\delta \Pi$ near the peak of the dome (at $n_e=1\times 10^{20}$ cm$^{-3}$), 
where $g_{TO}\approx 45.3$ meV and $k_Fa \approx 0.56$, which leads to $\sqrt{\delta\Pi}/k_F \sim 2.5$. Therefore, we conclude that the Landau damping term $\delta \Pi$ can not be neglected in FE doped STO for densities near the peak of the dome or above it. 

\subsection{\label{app3c} Gap equation for the generalized Rashba coupling}

In this appendix, we focus on the generalized Rashba coupling that accounts for an additional correction to the linear coupling from the nonlinear coupling. The coupling is described by the Lagrangian
\begin{equation}
 \begin{aligned}
L_{Rashba} &=\left(\frac{a}{L}\right)^3  \sum_{\bm{k},\bm{q}}  \psi_\alpha^\dagger\left(\hat{k}+\frac{\bm{q}}{2}\right) \bm{\eta}(\bm{q})  \cdot    \left(    \hat{k} \times \bm{\sigma}_{\alpha \gamma}    \right)  \left[ g \, \delta_{\gamma \beta}  +g_{NL} \, \bm{\eta}(\bm{q})  \cdot  \left(    \hat{k} \times \bm{\sigma}_{ \gamma \beta}    \right)        \right]   \psi_\beta \left(\bm{k}-\frac{\bm{q}}{2}\right) \\
&=\left(\frac{a}{L}\right)^3  \sum_{\bm{k},\bm{q}} g \psi^\dagger    \left(\bm{k}+\frac{\bm{q}}{2}\right)   \left[  \bm{\sigma} \cdot \left( \bm{\eta} \times \hat{k}    \right) \right] \psi \left(\bm{k}-\frac{\bm{q}}{2}\right) + \left(\frac{a}{L}\right)^3 g_{NL} \sum_{\bm{k},\bm{q}} \psi^\dagger    \left(\bm{k}+\frac{\bm{q}}{2}\right)    \bm{\eta} \cdot \hat{P}(\hat{k}) \cdot \bm{\eta} \,  \psi \left(\bm{k}-\frac{\bm{q}}{2}\right)\,.
	 \label{eq:e-ph_vec_quad_coulping_simpl}
 \end{aligned}
\end{equation}
Here, the first term represents the linear Rashba coupling, where the coupling constant is denoted by $g$.  The second term captures the quadratic coupling with a coupling constant, $g_{NL}$. The transverse displacement vector is denoted by $\bm{\eta}$.
In the second step, we have dropped the spin indices for simplicity and used the identity  $(\bm{\sigma}\cdot \bm{a}) (\bm{\sigma} \cdot \bm{b})=(\bm{a}\cdot \bm{b})\sigma_0+(\bm{a} \times \bm{b}) \bm{\sigma}$.

The equation for the pairing vertex associated with this coupling can be written as
\begin{equation}
 \begin{aligned}
	 \Phi_{\alpha\beta}(k) &=  \bar{g}  \frac{T}{ k_a^3}    \sum_{p_0}\int\frac{d^3p}{(2\pi)^3}  G(p)G(-p)D(p-k)    \bm{\gamma}^{ \nu \alpha}(-\hat{k})\Phi_{\nu \mu}(k)  \cdot \hat{P}(\hat{q}) \cdot \bm{\gamma}^{ \mu \beta}(\hat{k}) \\
	 & + 4\bar{g}_{NL}\langle \bm{\eta} \rangle^2  \frac{T}{ k_a^3}   \sum_{p_0}\int\frac{d^3p}{(2\pi)^3}  G(p)G(-p)D(p-k) \,     \bm{\gamma}^{ \tau \alpha}(-\hat{k}) \bm{\gamma}^{ \gamma \tau}(-\hat{k}) \cdot \hat{P}(\hat{q}) \Phi_{\gamma \mu}(k) \cdot \hat{P}(\hat{q}) \cdot   \bm{\gamma}^{ \mu  \nu}(\hat{k}) \bm{\gamma}^{  \nu \beta}(\hat{k})\,.
 \label{eq:vertex_tot1}
 \end{aligned}
\end{equation} 
Here, we have defined the coupling constants as $ \bar{g}=g^2 D_0 $, and $ \bar{g}_{NL}=g_{NL}^2  D_0 $ for linear and nonlinear coupling, respectively. 
We consider the attractive singlet channel. The spin summation and projection on the transverse sector for the linear coupling (the first term in Eq.~\eqref{eq:vertex_tot1}) is already discussed in Appendix~\ref{appA2} (see Eq.~\eqref{eq:vertex_vector1}). We proceed with the quadratic coupling term (the second term in Eq.~\eqref{eq:vertex_tot1}) in a similar fashion which leads to  
\begin{equation}
 \begin{aligned}
 & \quad   4 \langle \bm{\eta} \rangle^2  \left( -\hat{k} \times \bm{\sigma}  \right)_{ \tau \alpha}  \left( -\hat{k} \times \bm{\sigma}  \right)_{ \gamma \tau} \cdot \hat{P}(\hat{q}) \Phi_{\gamma \mu}(p_0,\bm{k}) \cdot  \hat{P}(\hat{q}) \cdot \left( \hat{k} \times \bm{\sigma}  \right)_{ \mu \nu} \left( \hat{k} \times \bm{\sigma}  \right)_{ \nu \beta} \\
 & =         4   \langle \bm{\eta} \rangle^2  i \sigma_y   \left( \sigma_t \sigma_t+ \sigma_u \sigma_u   \right)  \phi(p_0)  \left(   \cos \phi \sigma_t- \sin \phi \sigma_u   \right)  \left(   \cos \phi \sigma_t- \sin \phi \sigma_u   \right) \\
 &=8 \langle \bm{\eta} \rangle^2  i \sigma_y \phi(p_0) \, .
 \label{eq:vertex_term2}
 \end{aligned}
\end{equation}
Eq.~\eqref{eq:vertex_tot1} can now be simplified as
\begin{equation}
 \begin{aligned}
	 \phi(k_0) &=\frac{\bar{g}+8\bar{g}_{NL} \langle \bm{\eta} \rangle^2}{ k_a^3} T \sum_{p_0} \int\frac{d^3p}{(2\pi)^3}    G(p) \phi(p_0) G(-p)D(p-k) \\
  &=(\lambda+\lambda_{NL}  \langle \bm{\eta} \rangle^2)        T \sum_{p_0} \frac{ d(p_0-k_0)\phi(p_0)}{\vert p_0+\Sigma(p_0)\vert} \, .
 \label{eq:vertex_tot2_2}
 \end{aligned}
\end{equation}
Here, the effective coupling constant for the linear and nonlinear couplings are represented by $\lambda=\bar{g} \nu_F k_a^2/4k_F^2$ and $\lambda_{NL}  \langle \bm{\eta} \rangle^2=2 \bar{g}_{NL} \nu_F k_a^2 \langle \bm{\eta} \rangle^2/k_F^2$, respectively.  
We exclude the diverging contribution of $d(p_0-k_0)$ at $k_0=p_0$ and define a new gap function $\Delta(k_0)=k_0 \, \phi(k_0)/\tilde{\Sigma}(k_0)$ (similar to the derivation in Appendix~\ref{appA3}), which results in 
\begin{equation}
 \begin{aligned}
	&\Delta(k_0) =  \frac{(\lambda+\lambda_{NL}  \langle \bm{\eta} \rangle^2) T \sum_{p_0 \ne k_0} \frac{\Delta(p_0)}{p_0} \left[ d(p_0-k_0)+  d(p_0+k_0)  \right]}{1+(\lambda+\lambda_{NL}  \langle \bm{\eta} \rangle^2) T\frac{1}{k_0}\sum_{q_0 \ne k_0}  \left[ d(q_0-k_0)-  d(q_0+k_0)  \right] }\, .
	 \label{eq:eliashberg_gapeq_matrix_vecquad}
 \end{aligned}
\end{equation}
We solve this equation to calculate $T_c$. The results are discussed in Section~\ref{sec4}.

\section{\label{app4} Additional information on the numerical computation and the theoretical results}
\subsection{\label{app4a} The phase diagram} 
The calculated $T_c$, obtained from Eq.~\eqref{eq:eliashberg_gapeq_matrix} (associated with the linear coupling) and Eq.~\eqref{eq:eliashberg_gapeq_matrix_vecquad} (associated with the enhanced linear coupling from the nonlinear correction), is shown in panel (a) and panel (b) of Fig.~\ref{app_dometc_g_lin_quad}, respectively. In both panels, $T_c$ is plotted as a function of $-\omega_0^2$ and carrier density, $n_e$, with $n_e$ shown on a logarithmic scale. The results in panel (a) and panel (b) are identical to those in Fig.~\ref{dometc_g_DFT}(b) and Fig.~\ref{quad_lambda_Tc}(b), respectively, where carrier density, $n_e$, is presented on a linear scale.
\begin{figure}[t]
\includegraphics[width=\hsize]{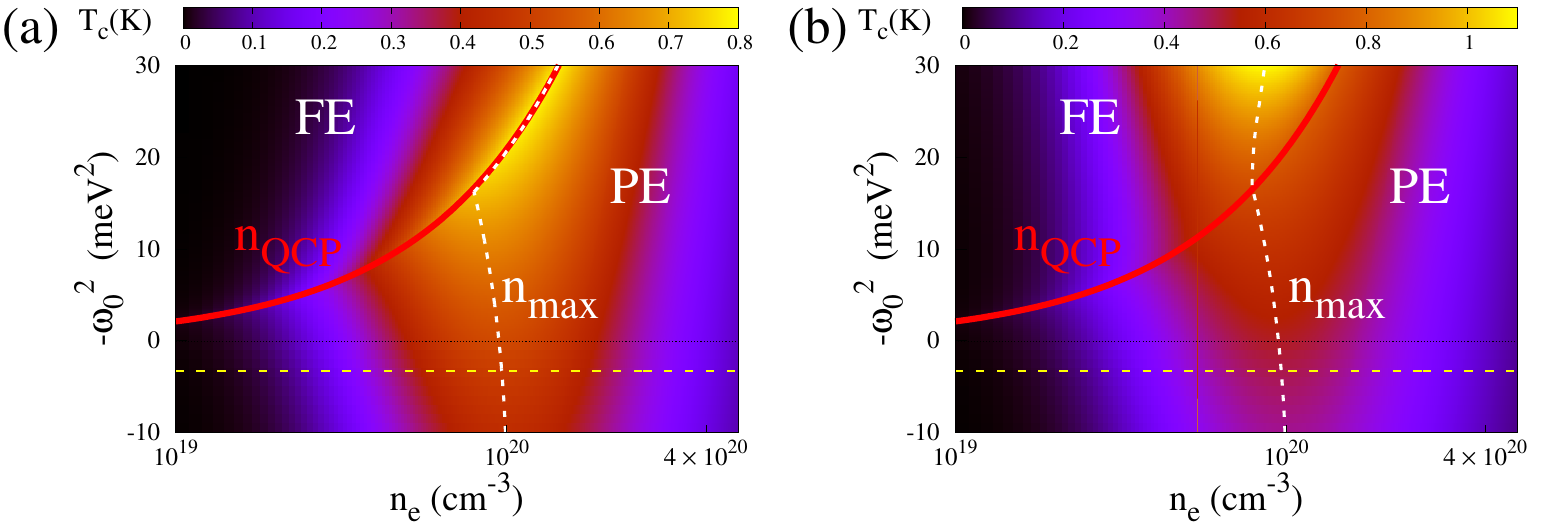}
\caption{\label{app_dometc_g_lin_quad} 
The pairing phase diagram as a function of carrier density $n_e$ and frequency $\omega_0^2$. Panel (a) corresponds to the linear coupling and panel (b) corresponds to the effective enhancement of the linear coupling from nonlinear contribution ($b=49$ meV$^2$). The yellow dashed line in both panels marks STO without FE doping ($\omega_{0}^2=3.3$ meV$^2$). The red solid line in both panels denotes $n_{QCP}$ and the white dashed line indicates the peak of the superconducting dome, $n_{max}$.}
\end{figure}

\subsection{\label{app4b}  Truncated linear coupling} 
In Section~\ref{sec3}, we demonstrated that the QC Eliashberg theory with linear coupling produces a peak in the superconducting dome at the QCP, resulting in a cusp at $n_{max}=n_{QCP}$, given that $\omega_0^2$ ($<0$) is sufficiently large to place the QCP in the high density regime where the coupling, $\lambda$, remains density independent. In this section, we show that the maximum $T_c$ can shift away from the QCP into the ordered phase when the coupling is truncated at large density.  
We incorporate a density cutoff in the linear coupling,  $g_{TO}$, for density $n_e>n_e^m$ ($n_e^m$ is the density where $g_{TO}$ is maximum) with a ``Fermi function''. The modified bare coupling $g_{TO}^c$ is given by
\begin{equation}
 \begin{aligned}
g_{TO}^c=g_{TO} \left[  \frac{1}{ 1+\exp\left( \frac{n_e- \alpha \, n_e^{m}}{\beta \, 10^{20}}  \right) } \right]\, .
 \label{eq:gto_modify}
  \end{aligned}
 \end{equation}
The purpose of using the ``Fermi function'' is to truncate $g_{TO}$ in the high density regime, while preserving its behavior in the low density regime. The parameters $\alpha$ and $\beta$ govern the range of these two regimes along with the decay rate at large density.
The modified couplings, $g_{TO}^c$ and $\lambda^c=(g_{TO}^c)^2 D_0\nu_F$ with $\alpha=4$ and $\beta=0.2$ are shown in the panel (a) of Fig.~\ref{gc_Tc} by the dashed lines. These parameter values indicate a strong decay chosen to enable a prominent comparison of $T_c$ with the previous results.
The truncated coupling, $\lambda^c$, increases with density in the low density regime and decreases in the high density regime. In the intermediate regime, around $n_e=0.9 \times 10^{20}$ cm$^{-3}$, $\lambda^c$ exhibits a weak density dependence, forming a rounded top.  This behavior arises as an artifact of the ``Fermi function''. 

\begin{figure}[h]
\includegraphics[width=\hsize]{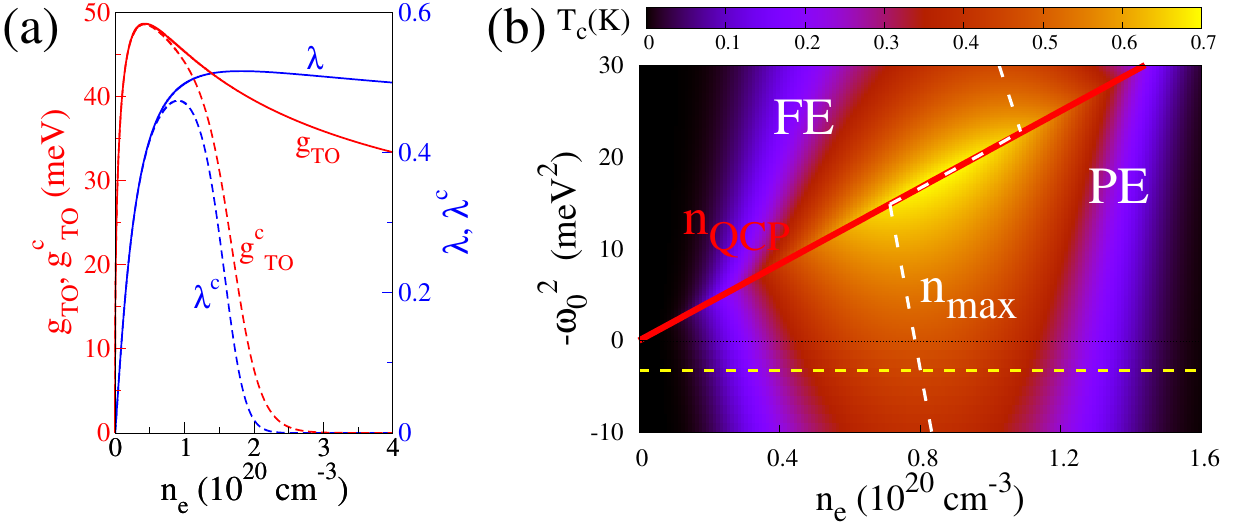}
\caption{\label{gc_Tc} Panel (a): The red lines represent the bare electron-phonon coupling (left y-axis) and the blue lines denote the effective coupling constant (right y-axis) as a function of carrier density $n_e$. The solid lines represent the linear coupling $g_{TO}$ ~\cite{maria2} (see Eq.~\eqref{eq:param_gto}) and $\lambda=g_{TO}^2 D_0\nu_F$ and the dashed lines denote the truncated couplings $g_{TO}^c$ (see Eq.~\eqref{eq:gto_modify}) and $\lambda^c=(g_{TO}^c)^2 D_0\nu_F$ with $\alpha=4$ and $\beta=0.2$.
Panel (b): The color map represents $T_c$ as a function of $n_e$ and $-\omega_0^2$ along $x$-axis and $y$-axis, respectively. The results correspond to the truncated coupling $g_{TO}^c$ with $\alpha=4$ and $\beta=0.2$. Additionally, $g_{TO}^c$ is rescaled ($g_{TO}^c\rightarrow g_{TO}^c/1.1$) such that the maximum $T_c$ for $\omega_{O}^2=3.3$ agrees with the experimentally measured value of the same for STO without FE doping. The yellow dashed line marks STO without FE doping ($\omega_{0}^2=3.3$ meV$^2$). The red solid line denotes $n_{QCP}$ and the white dashed line shows the peak position of the SC dome, $n_{max}$.}
\end{figure}

We solve the gap equation in Eq.~\eqref{eq:eliashberg_gapeq_matrix} by replacing $\lambda$ with the truncated effective coupling $\lambda^c$. The resulting $T_c$ is shown by the color map in panel (b) of Fig.~\ref{gc_Tc} as a function of carrier density, $n_e$, and $\omega_0^2$. 
The position of the dome peak, $n_{max}$, is indicated by the white dashed line and the red solid line represents $n_{QCP}$.
The figure shows that the dome peak shifts to the lower density as $\omega_0^2$ decreases (when $\omega_0^2$ is small, whether positive or negative). At a certain value of $\omega_0^2$ ($<0$), $n_{max}$ reaches the QCP. The dome peak coincides with the QCP for a small range of $\omega_0^2$ ($<0$).  This regime corresponds to the weak density dependence of the truncated coupling $\lambda^c$ in the intermediate density regime.
As $\omega_0^2$ decreases further, the dome peak shifts into the ordered state. However, $\lambda^c$ at $n_{QCP}$ is significantly suppressed in this regime, resulting in reduced values of $T_c$.

\subsection{\label{app4c} Rescaled frequency}

We have presented the gap equations for the linear coupling and the enhanced linear coupling, incorporating the nonlinear correction, in Eq.~\eqref{eq:eliashberg_gapeq_matrix} and Eq.~\eqref{eq:eliashberg_gapeq_matrix_vecquad}, respectively.
In both the expressions, the upper cutoff of the summations over fermionic frequency $p_0=(2m+1)\pi T$ is fixed at the Fermi level $f_c=E_F/k_B$. The frequency index $m$ is an integer number. For a fixed maximum frequency index $M$ we redefine the energy scale of frequencies with index $m>m_{rg}$, while keeping the regular energy scale for index $m<m_{rg}$ ~\cite{freqsc}. The rescaled frequencies can be presented as
\begin{equation}
 \begin{aligned}
p_0 &=(2m+1) \pi T, \qquad \qquad \qquad \qquad \quad m < m_{rg} \\
p_0 &= (2m+1) \pi T + 2 e^{am-b} \pi T \qquad \qquad m_{rg} \le m \le M \\
&= \left[2(m+e^{am-b}) +1 \right]\pi T \, .
\label{eq:frq_scale}
 \end{aligned}
 \end{equation}
These expressions suggest that the modified frequency indices for $m>m_{rg}$ are given by $m_{eff}=m+e^{am-b}$. The rescaled frequency indices are separated from each other by
\begin{equation}
 \begin{aligned}
df(m)&=1 \,,  \qquad \qquad \qquad \qquad \qquad \quad m < m_{rg} \\
&=1+e^{am-b}(e^a-1)\, . \qquad \qquad m_{rg} \le m \le M
\label{eq:frq_gap}
 \end{aligned}
 \end{equation}
The parameters, $a$ and $b$, are chosen so that the deviation from the regular frequency scale at $m=m_{rg}$ is tiny ($s$). This deviation gradually increases with higher indices, eventually reaching the desired upper cutoff frequency, $f_c$, at $m=M$. Additionally, the upper limit $f_c$ can be defined in terms of the regular frequency indices as $f_c=(2k+1)\pi T$. Thus, Eq.~\eqref{eq:frq_scale} leads to
\begin{equation}
 \begin{aligned}
&m_{rg}+e^{am_{rg}-b}=m_{rg}+s \, , \\
& am_{rg}-b=\ln{s} \, .
\label{eq:frq_scale_c1}
 \end{aligned}
 \end{equation}
\begin{equation}
 \begin{aligned}
&M+e^{aM-b}=k \, , \\
& aM-b=\ln{(k-M)} \, .
\label{eq:frq_scale_c2}
 \end{aligned}
 \end{equation}
Now, from Eq.~\eqref{eq:frq_scale_c1}, and ~\ref{eq:frq_scale_c2}, we obtain 
\begin{equation}
 \begin{aligned}
&a=\frac{\ln{(k-M)}-\ln{s}}{M-m_{rg}} \,, \\
&b=\frac{m_{rg}\ln{(k-M)}-M\ln{s}}{M-m_{rg}}\,.
\label{eq:frq_scale_c3}
 \end{aligned}
 \end{equation}
In Fig.~\ref{frq_rescale}, we plot $m_{eff}$ as a function of $m$ with three different values of $M$ ($M=20$, $50$ and $100$). The parameters used are $m_{rg}=10$, frequency shift $s=0.1$ from the regular frequency at $m=m_{rg}$, and $k=1000$. The figure demonstrates how the $M$ frequency indices effectively span the entire energy range up to $f_c$.
\begin{figure}[t]
\includegraphics[width=0.5\hsize]{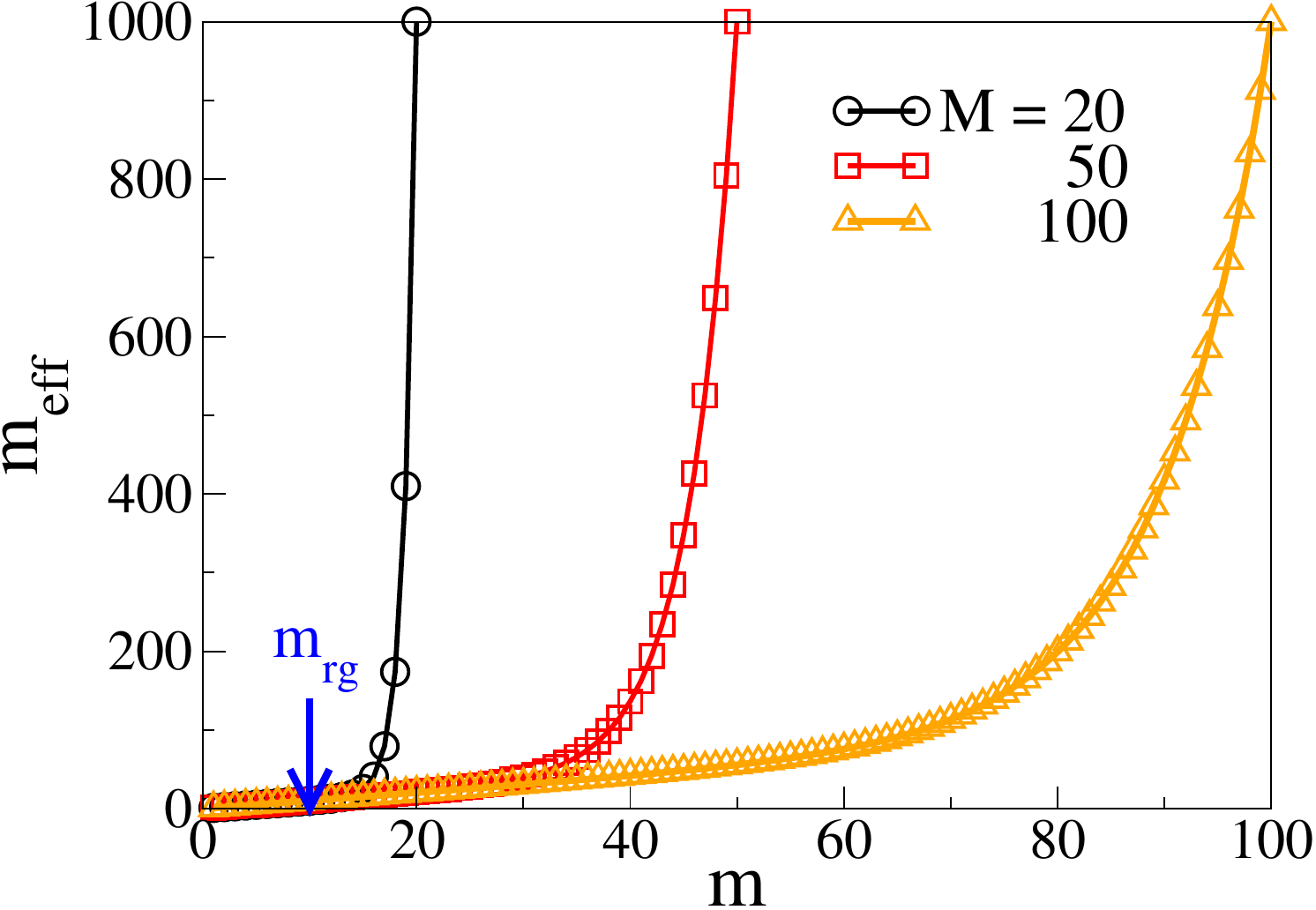}
\caption{\label{frq_rescale}  Rescaled frequency index $m_{eff}$ grows to $k=1000$ at $m=M$. We choose three different values of $M$ as $M=20$, $50$, $100$. The highest index cutoff of the regular frequency scale in the rescaled frequency, $m_{rg}=10$, is marked by the arrow. }
\end{figure}

Finally, let's assume that the indices for frequencies $p_0$ and $q_0$ in Eq.~\eqref{eq:eliashberg_gapeq_matrix} and Eq.~\eqref{eq:eliashberg_gapeq_matrix_vecquad} are $m$ and $n$, respectively. We use the rescaled frequencies in these expressions. Thus, Eq.~\eqref{eq:eliashberg_gapeq_matrix} is modified as
\begin{equation}
 \begin{aligned}
	\Delta(k_0)=\frac{\lambda T \sum_{p_0 \ne k_0} df(m) \frac{\Delta(p_0)}{p_0} \left[ d(p_0-k_0)+  d(p_0+k_0)  \right]}{1+\lambda T\frac{1}{k_0}\sum_{q_0 \ne k_0}  df(n)\left[ d(q_0-k_0)-  d(q_0+k_0)  \right] } \, .
	 \label{eq:eliashberg_gapeq_matrix_mn_resc}
 \end{aligned}
\end{equation}
Similarly, Eq.~\eqref{eq:eliashberg_gapeq_matrix_vecquad} is modified as
\begin{equation}
 \begin{aligned}
	\Delta(k_0) =  \frac{(\lambda+\lambda_{NL}  \langle \bm{\eta} \rangle^2) T \sum_{p_0 \ne k_0} df(m) \frac{\Delta(p_0)}{p_0} \left[ d(p_0-k_0)+  d(p_0+k_0)  \right]}{1+(\lambda+\lambda_{NL}  \langle \bm{\eta} \rangle^2) T\frac{1}{k_0}\sum_{q_0 \ne k_0} df(n)  \left[ d(q_0-k_0)-  d(q_0+k_0)  \right] } \, .
	 \label{eq:eliashberg_gapeq_matrix_vecquad_mn_resc}
 \end{aligned}
\end{equation}
\twocolumngrid
\bibliography{ferroelectric_ref}

\end{document}